\documentclass[english,traditabstract]{aa}
\usepackage[varg]{txfonts}
\usepackage[utf8]{inputenc}
\usepackage[T1]{fontenc}
\usepackage{prettyref}
\usepackage{array}
\usepackage{rotating}
\usepackage{float}
\usepackage{multirow}
\usepackage{amstext}
\usepackage{graphicx}
\usepackage{epstopdf}
\usepackage{subfigure}
\usepackage{gensymb}
\usepackage{longtable}
\makeatletter

\providecommand{\tabularnewline}{\\}

\usepackage{natbib,twoopt}
\usepackage[breaklinks=true]{hyperref} 
\usepackage{arydshln}
\bibpunct{(}{)}{;}{a}{}{,} 
\makeatletter
\newcommandtwoopt{\citeads}[3][][]{\href{http://adsabs.harvard.edu/abs/#3}%
{\def\hyper@linkstart##1##2{}%
\let\hyper@linkend\@empty\citealp[#1][#2]{#3}}}
\newcommandtwoopt{\citepads}[3][][]{\href{http://adsabs.harvard.edu/abs/#3}%
{\def\hyper@linkstart##1##2{}%
\let\hyper@linkend\@empty\citep[#1][#2]{#3}}}
\newcommandtwoopt{\citetads}[3][][]{\href{http://adsabs.harvard.edu/abs/#3}%
{\def\hyper@linkstart##1##2{}%
\let\hyper@linkend\@empty\citet[#1][#2]{#3}}}
\newcommandtwoopt{\citeyearads}[3][][]%
{\href{http://adsabs.harvard.edu/abs/#3}
{\def\hyper@linkstart##1##2{}%
\let\hyper@linkend\@empty\citeyear[#1][#2]{#3}}}
\makeatother
\titlerunning{Astrometric Positions of the Irregular Satellites of Giant Planets}
\authorrunning{Gomes-Júnior et al.}

\makeatother

\begin{document}

\title{Astrometric positions for 18 irregular satellites of giant planets from 23 years of observations
\fnmsep\thanks{Table 8 are only available in electronic form
at the CDS via anonymous ftp to cdsarc.u-strasbg.fr (130.79.128.5)
or via http://cdsweb.u-strasbg.fr/cgi-bin/qcat?J/A+A/ and IAU NSDC data base at www.imcce.fr/nsdc.}
\fnmsep\thanks{Partially based on observations made at Laborat\'orio Nacional de Astrof\'{\i}sica (LNA), Itajub\'a-MG, Brazil.}
\fnmsep\thanks{Partially based on observations through the ESO runs 079.A-9202(A), 075.C-0154, 077.C-0283 and 079.C-0345.}
\fnmsep\thanks{Partially based on observations made at
Observatoire de Haute Provence (OHP), F-04870 Saint-Michel l'observatoire, France}
}

\author{ A. R. Gomes-Júnior
          \inst{1},
          M. Assafin 
          \inst{1} \fnmsep\thanks{Affiliated researcher at Observatoire de Paris/IMCCE, 77 Avenue Denfert Rochereau 75014 Paris, France},
          R. Vieira-Martins
          \inst{1,2,3} \fnmsep\thanks{Affiliated researcher at Observatoire de Paris/IMCCE, 77 Avenue Denfert Rochereau 75014 Paris, France},
          J.-E. Arlot
          \inst{4},      
          J. I. B. Camargo
          \inst{2,3},     
          F. Braga-Ribas
          \inst{2,5}, 
          D. N. da Silva Neto
          \inst{6},
          A. H. Andrei
          \inst{1,2} \fnmsep\thanks{Affiliated researcher at Observatoire de Paris/SYRTE, 77 Avenue Denfert Rochereau, 75014 Paris, France},
          A. Dias-Oliveira
          \inst{2},
          B. E. Morgado
          \inst{1},
          G. Benedetti-Rossi
          \inst{2},
          Y. Duchemin
          \inst{4,7},
          J. Desmars
          \inst{4},
          V. Lainey
          \inst{4},
          W. Thuillot
          \inst{4}
          }

\offprints{A. R. Gomes-Júnior}

   \institute{Observatório do Valongo/UFRJ, Ladeira Pedro Antônio 43,
CEP 20.080-090 Rio de Janeiro - RJ, Brazil\\
              \email{altair08@astro.ufrj.br}
              \and
Observatório Nacional/MCT, R. General José Cristino 77,
CEP 20921-400 Rio de Janeiro - RJ, Brazil\\
              \email{rvm@on.br}
              \and
Laboratório Interinstitucional de e-Astronomia - LIneA, Rua Gal. José Cristino 77, Rio de Janeiro, RJ 20921-400, Brazil
              \and
Institut de mécanique céleste et de calcul des éphémérides - Observatoire de Paris, UMR 8028 du CNRS,
77 Av. Denfert-Rochereau, 75014 Paris, France\\
              \email{arlot@imcce.fr}
              \and
Federal University of Technology - Paraná (UTFPR / DAFIS), Rua Sete de Setembro, 3165, CEP 80230-901, Curitiba, PR, Brazil
              \and
Centro Universitário Estadual da Zona Oeste, Av. Manual Caldeira de Alvarenga 1203, CEP 23.070-200 Rio de Janeiro RJ, Brazil
              \and
              ESIGELEC-IRSEEM, Technopôle du Madrillet, Avenue Galilée, 76801 Saint-Etienne du Rouvray, France
              }

\date{Received: Abr 08, 2015; accepted: May 06, 2015}

\abstract
{The irregular satellites of the giant planets are believed to have been captured during the evolution of the solar system. Knowing their physical parameters, such as size, density, and albedo is important for constraining where they came from and how they were captured. The best way to obtain these parameters are observations in situ by spacecrafts or from stellar occultations by the objects. Both techniques demand that the orbits are well known.}
{We aimed to obtain good astrometric positions of irregular satellites to improve their orbits and ephemeris.}
{We identified and reduced observations of several irregular satellites from three databases containing more than 8000 images obtained between 1992 and 2014 at three sites (Observatório do Pico dos Dias, Observatoire de Haute-Provence, and European Southern Observatory - La Silla).  We used the software PRAIA (Platform for Reduction of Astronomical Images Automatically) to make the astrometric reduction of the CCD frames. The UCAC4 catalog represented the International Celestial Reference System in the reductions. Identification of the satellites in the frames was done through their ephemerides as determined from the SPICE/NAIF kernels. Some procedures were followed to overcome missing or incomplete information (coordinates, date), mostly for the older images.}
{We managed to obtain more than 6000 positions for 18 irregular satellites: 12 of Jupiter, 4 of Saturn, 1 of Uranus (Sycorax), and 1 of Neptune (Nereid). For some satellites the number of obtained positions is more than 50\% of what was used in earlier orbital numerical integrations.}
{Comparison of our positions with recent JPL ephemeris suggests there are systematic errors in the orbits for some of the irregular satellites. The most evident case was an error in the inclination of Carme.}


\maketitle

\section{Introduction} \label{Sec: introducao} 

The irregular satellites of the giant planets are smaller than the regular moons, having more eccentric, inclined, distant, and in most cases, retrograde orbits. Owing to their orbital configurations, it is largely accepted that these objects were captured in the early solar system \citep{Sheppard2003}.

Because they are faint, the majority of these objects were only discovered in the last century\footnote{Website: http://ssd.jpl.nasa.gov/?sat\_discovery} . They were never visited by a spacecraft, with the exception of Himalia, Phoebe and Nereid, in a flyby by the Cassini space probe in 2000 for Himalia \citep{Porco2003} and in 2004 for Phoebe \citep{Desmars2013} and in a flyby by the Voyager 2 space probe in 1989 for Nereid \citep{Smith1989}. Even in situ, they were still opportunity target observations resulting in not optimal measurements, with size errors of $10 km$ for Himalia and $25 km$ for Nereid \citep{Thomas1991}. The exception is Phoebe with a very accurate measurement of size with a mean radius error of $0.7 km$ \citep{Thomas2010}.

If these objects were captured, there remains the question of where they came from. \citealp{Clark2005} show from imaging spectroscopy from Cassini that Phoebe has a surface probably covered by material from the outer solar system and \citealp{Grav2003} show that the satellites of the Jovian Prograde Group Himalia have gray colors implying that their surfaces are similar to that of C-type asteroids. In that same work, the Jovian Retrograde Group Carme was found to have surface colors similar to the D-type asteroids as for the Hilda or Trojan families, while JXIII Kalyke has a redder color like Centaurs or trans-Neptunian objects (TNOs).

For Saturnian satellites, \citealp{Grav2007} show by their colors and spectral slopes that these satellites contain a more or less equal fraction of C-, P-, and D-like objects, but SXXII Ijiraq is marginally redder than D-type objects. These works may suggest different origins for the irregular satellites.

In this context, we used three databases for deriving precise positions for the irregular satellites observed at the Observatório do Pico dos Dias (1.6 m and 0.6 m telescopes, IAU code 874), the Observatoire Haute-Provence (1.2m telescope, IAU code 511), and ESO (2.2 m telescope, IAU code 809). Many irregular satellites were observed between 1992 and 2014, covering a few orbital periods of these objects (12 satellites of Jupiter, 4 of Saturn, Sycorax of Uranus, and Nereid of Neptune). 

Since their ephemerides are not very precise, predicting and observing stellar occultations are very difficult, and no observation of such an event for an irregular satellite is found in the literature. The precise star positions to be derived by the ESA astrometry satellite Gaia \citep{deBruijne2012} will render better predictions with the only source of error being the ephemeris. The positions derived from our observations can be used in new orbital numerical integrations, generating more precise ephemerides.

The power of stellar occultations for observing relatively small diameter solar system objects is supported by recent works, such as the discovery of a ring system around the Centaur (10199) Chariklo \citep{Braga-Ribas2014}. Once irregular satellites start to be observed by this technique, it will be possible to obtain their physical parameters (shape, size, albedo, density) with unprecedented precision. For instance, in this case, sizes could be obtained with kilometer accuracy. The knowledge of these parameters would in turn bring valuable information for studying the capture mechanisms and origin of the irregular satellites.

The databases are described in Sect. \ref{Sec: observations}. The astrometric procedures in Sect. \ref{Sec: reduction}. The obtained positions are presented in Sect \ref{Sec: positions} and analyzed in Sect. \ref{Sec: comparison}. Conclusions are given in Sect. \ref{Sec: conclusions}.

\section{Databases} \label{Sec: observations}

Our three databases consist of optical CCD images from many observational programs performed with different telescopes and detectors that target a variety of objects, among which are irregular satellites. The observations were made at three sites: Observatório do Pico dos Dias (OPD), Observatoire Haute-Provence (OHP), and the European Southern Observatory (ESO). All together there are more than 8000 FITS images obtained in a large time span (1992-2014) for the irregular satellites. Since the OHP and mostly the OPD database registers were not well organized, we had to start from scratch and develop an automatic procedure to identify and filter only the images of interest, that is, for the irregular satellites. The instrument and image characteristics are described in the following sections.

\subsection{OPD} \label{Subsec: observations-opd}

The OPD database was produced at the Observatório do Pico dos Dias (OPD, IAU code 874, $45^{\circ} ~34\arcmin ~57\arcsec$ W, $22^{\circ} ~32\arcmin ~04\arcsec$ S, 1864 m)\footnote{Website: http://www.lna.br/opd/opd.html - in Portuguese}, located at geographical longitude , in Brazil. The observations were made between 1992 and 2014 by our group in a variety of observational programs. 
Two telescopes of 0.6 m diameter (Zeiss and Boller \& Chivens) and one 1.6 m diameter (Perkin-Elmer) were used for the observations. Identified were 5248 observations containing irregular satellites, with 3168 from the Boller \& Chivens, 1967 from the Perkin-Elmer, and 113 from the Zeiss.


This is an inhomogeneous database with observations made with nine different detectors (see Table \ref{Tab: OPD-CCDs}) and six different filters. The headers of most of the older FITS images had missing, incomplete, or incorrect coordinates or dates. In some cases, we could not identify the detector's origin. The procedures used to overcome these problems are described in Sect. \ref{Sec: reduction}.

\begin{table}[h!]
\caption{\label{Tab: OPD-CCDs} Characteristics of OPD detectors used in this work.}
\begin{centering}
\begin{tabular}{ccc}
\hline
\hline
\multicolumn{3}{c}{Perkin-Elmer}\tabularnewline
Detector & Image size (pixel) & Pixel Scale ($\mu m$/px)\tabularnewline
\hline
CCD048 & 770 x 1152 & 22.5\tabularnewline
CCD098 & 2048 x 2048 & 13.5\tabularnewline
CCD101 & 1024 x 1024 & 24.0\tabularnewline
CCD105 & 2048 x 2048 & 13.5\tabularnewline
CCD106 & 1024 x 1024 & 24.0\tabularnewline
CCD301 & 385 x 578 & 22.0\tabularnewline
CCD523 & 455 x 512 & 19.0\tabularnewline
IKON & 2048 x 2048 & 13.5\tabularnewline
IXON & 1024 x 1024 & 13.5\tabularnewline
\hline 
\end{tabular} 
\par\end{centering}
The plate scale of the telescopes are 13.09"/mm for Perkin-Elmer, 25.09"/mm for Boller \& Chivens and 27.5"/mm for Zeiss.
\end{table}

%
%
%

\subsection{OHP} \label{Subsec: observations-ohp}

The instrument used at the Observatoire de Haute Provence (OHP, IAU code 511, $5^{\circ} ~42\arcmin ~56.5\arcsec$ E, $43^{\circ} ~55\arcmin ~54.7\arcsec $N, 633.9 m)\footnote{Website: www.obs-hp.fr/guide/t120.shtml - in French} was the 1.2m-telescope in a Newton configuration. The focal length is 7.2 m. The observations were made between 1997 and 2008. During this time only one CCD detector $1024 \times 1024$ was used. The size of field is $12\arcmin \times 12\arcmin$ with a pixel scale of $0.69\arcsec$.  
From these observations, 2408 were identified containing irregular satellites.

%
%
%
%
%

\subsection{ESO} \label{Subsec: observations-eso}

Observations were made at the 2.2 m Max-Planck ESO (ESO2p2) telescope (IAU code 809, $70\degr44\arcmin1.5\arcsec$ W, $29\degr15\arcmin31.8\arcsec$ S, 2345.4 m)\footnote{Website: www.eso.org/sci/facilities/lasilla/telescopes/\\national/2p2.html} with the Wide Field Imager (WFI) CCD mosaic detector. Each mosaic is composed of eight CCDs of $7.5\arcmin \times 15\arcmin$ ($\alpha$, $\delta$) sizes, resulting in a total coverage of $30\arcmin \times 30\arcmin$ per mosaic. Each CCD has $4 k \times 2 k$ pixels with a pixel scale of $0.238\arcsec$. The filter used was a broad-band R filter (ESO\#844) with $\lambda _{c}  = 651.725$ nm and $\Delta \lambda = 162.184$ nm. The telescope was shifted between exposures in such a way that each satellite was observed at least twice in different CCDs.

The satellites were observed in 24 nights, divided in five runs between April 2007 and May 2009 in parallel with, and using the same observational and astrometric procedures of the program that observed stars along the sky path of trans-Neptunian objects (TNOs) to identify candidates for stellar occultation \citep[see][]{2010A&A...515A..32A, 2012A&A...541A.142A}. A total of 810 observations were obtained for irregular satellites. 

%
%
%

\section{Astrometry} \label{Sec: reduction}

Almost all the frames were photometrically calibrated with auxiliary bias and flat-field frames by means of standard procedures using IRAF\footnote{Website: http://iraf.noao.edu/} and, for the mosaics, using the esowfi \citep{Jones2000} and mscred \citep{Valdes1998} packages. Some of the nights at OPD did not have bias and flat-field images so the correction was not possible.

The astrometric treatment was made with the Platform for Reduction of Astronomical Images Automatically (PRAIA) \citep{2011gfun.conf...85A}. The (x, y) measurements were performed with two-dimensional circular symmetric Gaussian fits within one full width half maximum (FWHM = seeing). Within one FWHM, the image profile is described well by a Gaussian profile, which is free of the wing distortions, which may jeopardize the determination of the center. PRAIA automatically recognizes catalog stars and determines ($\alpha$, $\delta$) with a user-defined model relating the (x, y) measured and (X, Y) standard coordinates projected in the sky tangent plane.

We used the UCAC4 \citep{2013AJ....145...44Z} as the practical representative of the International Celestial Reference System (ICRS). For each frame, we used the six constants polynomial model to relate the (x, y) measurements with the (X, Y) tangent plane coordinates. For ESO, we followed the same astrometric procedures as described in detail in \cite{2012A&A...541A.142A}; the (x, y) measurements of the individual CCDs were precorrected by a field distortion pattern, and all positions coming from different CCDs and mosaics were then combined using a third degree polynomial model to produce a global solution for each night and field observed, and final ($\alpha$, $\delta$) object positions were obtained in the UCAC4 system. 

In Table \ref{Tab: stars-errors} we list the average mean error in $\alpha$ and $\delta$ for the reference stars obtained by telescope, the average (x, y) measurement errors of the Gaussian fits described above, and the mean number of UCAC4 stars used by frame. For all databases, about 20\% of outlier reference stars were eliminated for presenting (O-C) position residuals higher than 120 mas in the ($\alpha$, $\delta$) reductions. 

\begin{table}
\caption{\label{Tab: stars-errors} Astrometric ($\alpha$, $\delta$) reduction by telescope.}
\begin{centering}
\begin{tabular}{lccccc}
\hline 
\hline
 & \multicolumn{2}{c}{Mean errors} &  UCAC4 & \multicolumn{2}{c}{Gaus. errors}       \tabularnewline
Telescope  & $\sigma_{\alpha}$  & $\sigma_{\delta}$  & stars & x & y  \tabularnewline
  &   mas   &   mas & &   mas   &   mas  \tabularnewline
\hline
PE(OPD) & 51 & 48 & 24 & 15 & 15 \tabularnewline
B\&C (OPD) & 56 & 55 & 36 & 29 & 29 \tabularnewline
Zeiss (OPD) & 58 & 57 & 95 & 26 & 26 \tabularnewline
OHP & 50 & 49 & 46 & 26 & 26 \tabularnewline
ESO & 26 & 25 & 632 & 15 & 15 \tabularnewline
\hline
\end{tabular}
\par\end{centering}
Mean errors are the standard deviations in the (O$-$C) residuals from ($\alpha$, $\delta$) reductions with the UCAC4 catalog. Gaussian errors are the errors in the Gaussian fit used to perform the (x, y) measurements.
\end{table}

To help identify the satellites in the frames and derive the ephemeris for the instants of the observations for comparisons (see Sect \ref{Sec: comparison}), we used the kernels from SPICE/JPL\footnote{Website: http://naif.jpl.nasa.gov/naif/toolkit.html}. \cite{Emelyanov2008} and references therein also provided ephemeris of similar quality for the irregular satellites. For instance, for Himalia, which has relatively good orbit solutions, the ephemerides differ by less than $20~mas$, and in the case of less-known orbits, like Ananke, the differences are less than $90~mas$. We chose to use the JPL ephemeris because they used more recent observations \cite[see][]{Jacobson2012}. The JPL ephemeris that represented the Jovian satellites in this work was the DE421 $+$ JUP300. For the Saturnian satellites, the ephemeris was DE421 $+$ SAT359 to Hyperion, Iapetus, and Phoebe and DE421 $+$ SAT361 to Albiorix, Siarnaq, and Paaliaq. The DE421 $+$ URA095 was used for Sycorax and DE421 $+$ NEP081 for Nereid. More recent JPL ephemeris versions became available after completion of this work, but this did not affect the results. 

In the OPD database, there were some images (mostly the older ones) with missing coordinates or the wrong date in their headers. In the case of missing or incorrect coordinates, we adopted the ephemeris as the central coordinates of the frames. When the time was not correct, the FOV identification failed. In this case, a search for displays of wrong date (year) was performed. Problems like registering local time instead of UTC were also identified and corrected.

In all databases, for each night a sigma-clipping procedure was performed to eliminate discrepant positions (outliers). A threshold of 120 mas and a deviation of more than 2.5 sigma from the nightly average ephemeris offsets were adopted.

From Tables \ref{Tab: Reductions-160} to \ref{Tab: Reductions-eso} we list the average dispersion (standard deviation) of the position offsets with regard to the ephemeris for $\alpha$ and $\delta$ obtained by telescope for each satellite. The final number of frames, number of nights (in parenthesis), the mean number of UCAC4 stars used in the reduction and the approximate V magnitude are also given. The dashed lines separate the satellites from different families with similar orbital parameters: Himalia Group (Himalia, Elara, Lysithea and Leda), Pasiphae Group (Pasiphae, Callirrhoe and Megaclite), and Ananke Group (Ananke and Praxidike). Carme and Sinope are the only samples of their groups. From Saturn, Siarnaq and Paaliaq are from the Inuit Group while Phoebe and Albiorix are the only samples in their groups.

The differences in the dispersion of the ephemeris offsets of the same satellite for distinct telescopes seen in Tables \ref{Tab: Reductions-160} to \ref{Tab: Reductions-eso} are caused by the different distribution of observations along the orbit for each telescope. This can be seen in Fig. \ref{Fig: carme_anom} for Carme and Fig. \ref{Fig: pasiphae_anom} for Pasiphae and for all objects in the online material. Since the observations cover different segments of the orbit, the dispersion of the offsets may vary for different telescopes for a single satellite, with larger covered segments usually implying larger dispersions and vice versa. For Nereid, due to its high eccentric orbit, the observations are located between 90$\degree$ and 270$\degree$ of true anomaly where Nereid remains most of the time.

\begin{table}
\caption{\label{Tab: Reductions-160} Astrometric ($\alpha$, $\delta$) reduction for each satellite observed with the Perkin-Elmer telescope.}
\begin{centering}
\begin{tabular}{lcccccc}
\hline 
\hline
 & \multicolumn{2}{c}{Offsets (sigma)} & Nr &  UCAC4 & \tabularnewline
Satellite  & $\sigma_{\alpha}$  & $\sigma_{\delta}$ & frames  & stars & Mag \tabularnewline
  &   mas   &   mas  & (nights)  & &\tabularnewline
\hline
Himalia & 290 & 45 & 238 (18) & 37 & 14 \tabularnewline
Elara & 230 & 118 & 99 (12) & 32 & 16\tabularnewline
Lysithea & 107 & 79 & 53 (8) & 41 & 18\tabularnewline
Leda & 207 & 79 & 6 (2) & 46 & 19\tabularnewline
\hdashline
Pasiphae & 157 & 92 & 144 (13) & 22 & 17 \tabularnewline
Callirrhoe & 66 & 35  &   9 (1) & 3 & 21\tabularnewline
\hdashline
Carme & 97 & 94 & 68 (7) & 49 & 18\tabularnewline
Sinope & 155 & 77 & 37 (8) & 42 & 18\tabularnewline
Ananke & 93 & 185 & 52 (7) & 40 & 19\tabularnewline
\hline
Phoebe & 73 & 95 & 410 (22) & 6 & 16\tabularnewline
\hline
Nereid & 200 & 142 & 289 (29) & 8 & 19\tabularnewline
\hline
\end{tabular}
\par\end{centering}
The offsets (sigma) are the average standard deviations of the ephemeris offsets from the ($\alpha$, $\delta$) positions of the satellites. Also given are the approximate satellite V magnitude and the average number of UCAC4 reference stars per frame.
\end{table}

\begin{table}
\caption{\label{Tab: Reductions-iag} Astrometric ($\alpha$, $\delta$) reduction for each satellite observed with the Boller \& Chivens telescope.}
\begin{centering}
\begin{tabular}{lccccc}
\hline 
\hline
 & \multicolumn{2}{c}{Offsets (sigma)} & Nr &  UCAC4 & \tabularnewline
Satellite  & $\sigma_{\alpha}$  & $\sigma_{\delta}$ & frames  & stars & Mag \tabularnewline
  &   mas   &   mas  & (nights)  & &\tabularnewline
\hline
Himalia & 83 & 43 & 560 (31) & 57 & 14 \tabularnewline
Elara & 55 & 43 & 294 (23) & 53 & 16 \tabularnewline
Lysithea & 23 & 42 & 7 (2) & 60 & 18 \tabularnewline
\hdashline
Pasiphae & 128 & 71 & 140 (14) & 57 & 17 \tabularnewline
Carme & 68 & 111 & 22 (4) & 45 & 18\tabularnewline
Sinope & 59 & 17  &   4 (1) & 22 & 18\tabularnewline
\hline
Phoebe & 43 & 48 & 810 (42) & 17 & 16 \tabularnewline
\hline
Nereid & 61 & 45 & 514 (38) & 20 & 19 \tabularnewline
\hline
\end{tabular}
\par\end{centering}
Same as in Table \ref{Tab: Reductions-160}.
\end{table}

\begin{table}
\caption{\label{Tab: Reductions-zei} Astrometric ($\alpha$, $\delta$) reduction for each satellite observed with the Zeiss telescope.}
\begin{centering}
\begin{tabular}{lccccc}
\hline 
\hline
 & \multicolumn{2}{c}{Offsets (sigma)} & Nr &  UCAC4 & \tabularnewline
Satellite  & $\sigma_{\alpha}$  & $\sigma_{\delta}$ & frames  & stars & Mag \tabularnewline
  &   mas   &   mas  & (nights)  & &\tabularnewline
\hline
Himalia & 112 & 72 & 56 (4) & 91 & 14\tabularnewline
Elara & 17 & 21  &  10 (1) & 146 & 16 \tabularnewline
\hdashline
Pasiphae & 24 & 25  &  11 (1) & 140 & 17\tabularnewline
\hline
Phoebe & 37 & 30  &  19 (1) & 16 & 16\tabularnewline
\hline
\end{tabular}
\par\end{centering}
Same as in Table \ref{Tab: Reductions-160}.
\end{table}

\begin{table}
\caption{\label{Tab: Reductions-ohp} Astrometric ($\alpha$, $\delta$) reduction for each satellite observed with the OHP telescope.}
\begin{centering}
\begin{tabular}{lccccc}
\hline
\hline
 & \multicolumn{2}{c}{Offsets (sigma)} & Nr &  UCAC4 & \tabularnewline
Satellite  & $\sigma_{\alpha}$  & $\sigma_{\delta}$ & frames  & stars & Mag \tabularnewline
  &   mas   &   mas  & (nights)  & &\tabularnewline
\hline
Himalia & 49 & 66 & 357 (43) & 49 & 14\tabularnewline
Elara & 52 & 61 & 187 (25) & 37 & 16\tabularnewline
Lysithea & 63 & 50 & 84 (13) & 56 & 18\tabularnewline
Leda & 118 & 33 & 48 (7) & 14 & 19\tabularnewline
\hdashline
Pasiphae & 101 & 75 & 248 (32) & 39 & 17\tabularnewline
Carme & 114 & 96 & 204 (29) & 39 & 18\tabularnewline
Sinope & 196 & 73 & 169 (25) & 43 & 18\tabularnewline
Ananke & 100 & 89 & 141 (20) & 62 & 19\tabularnewline
\hline
Phoebe & 30 & 31 & 516 (63) & 51 & 16\tabularnewline
Siarnaq & 46 & 98 & 20 (6) & 32 & 20\tabularnewline
\hline
\end{tabular}
\par\end{centering}
Same as in Table \ref{Tab: Reductions-160}.
\end{table}

\begin{table}
\caption{\label{Tab: Reductions-eso} Astrometric ($\alpha$, $\delta$) reduction for each satellite observed with the ESO telescope.}
\begin{centering}
\begin{tabular}{lccccc}
\hline 
\hline
 & \multicolumn{2}{c}{Offsets (sigma)} & Nr &  UCAC4 & \tabularnewline
Satellite  & $\sigma_{\alpha}$  & $\sigma_{\delta}$ & frames  & stars & Mag \tabularnewline
  &   mas   &   mas  & (nights)  & &\tabularnewline
\hline
Himalia & 76 & 74 & 23 (2) & 1153 & 14\tabularnewline
Elara & 112 & 87 & 46 (4) & 1492 & 16\tabularnewline
Lysithea & 76 & 88 & 90 (6) & 695 & 18\tabularnewline
Leda & 60 & 125 & 44 (3) & 632 & 19\tabularnewline
\hdashline
Pasiphae & 70 & 114 & 66 (5) & 836 & 17\tabularnewline
Callirrhoe & 29 & 33  &  16 (1) & 493 & 21\tabularnewline
Megaclite & 52 & 34  &  10 (1) & 445  & 22\tabularnewline
\hdashline
Ananke & 225 & 19 & 57 (3) & 761 & 18\tabularnewline
Praxidike & 7 & 38  &   2 (1) & 1934 & 21\tabularnewline
\hdashline
Carme & 140 & 110 & 37 (4) & 1074 & 18\tabularnewline
Sinope & 339 & 70 & 11 (2) & 1542 & 18\tabularnewline
Themisto & 894 & 28 & 16 (2) & 1232 & 21\tabularnewline
\hline
Phoebe & 102 & 57 & 32 (5) & 312 & 16\tabularnewline
\hdashline
Siarnaq & 86 & 66 & 56 (6) & 283 & 20\tabularnewline
Paaliaq & 301 & 59 & 11 (4) & 382 & 21\tabularnewline
\hdashline
Albiorix & 76 & 50 & 46 (6) & 330 & 20\tabularnewline
\hline
Sycorax & 150 & 82 & 35 (9) & 375 & 21\tabularnewline
\hline
Nereid & 115 & 78 & 99 (12) & 362 & 19\tabularnewline
\hline
\end{tabular}
\par\end{centering}
Same as in Table \ref{Tab: Reductions-160}.
\end{table}

No solar phase correction was applied to the positions. For the biggest irregular satellite of Jupiter, Himalia, it was verified that the maximum deviation in the position due to phase angle is 1.94 \textit{mas} using the phase correction described in \cite{Lindegren1977}. For the other satellites, which are smaller objects, this deviation is even smaller. Since our position error is one order of magnitude higher, this effect was neglected.

\section{Satellite positions} \label{Sec: positions}

The final set of positions of the satellites consists in 6523 cataloged positions observed between 1992 and 2014 for 12 satellites of Jupiter, 4 of Saturn, 1 of Uranus, and 1 of Neptune. The topocentric positions are in the ICRS. The catalogs (one for each satellite) contain epoch of observations, the position error, filter used, estimated magnitude (from PSF fitting) and telescope origin. The magnitude errors can be as high as 1 mag; they are not photometrically calibrated and should be used with care. The position errors were estimated from the dispersion of the ephemeris offsets of the night of observation of each position. Thus, these position errors are probably overestimated because there must be ephemeris errors present in the dispersion of the offsets. These position catalogs are freely available in electronic form at the CDS (see a sample in Table \ref{Tab: sample-cds}) and at the IAU NSDC data base at www.imcce.fr/nsdc.

The number of positions acquired is significant compared to the number used in the numerical integration of orbits by the JPL \citep{Jacobson2012} as shown in Table \ref{Tab: comparison-horizons}.

\begin{table*}
\caption{\label{Tab: sample-cds} CDS data table sample for Himalia.}
\begin{centering}
\begin{tabular}{ccccccccc}
\hline 
\hline
\multicolumn{2}{c}{RA  (ICRS)  Dec} & RA error & Dec error & Epoch & Mag & Filter & Telescope & IAU code \tabularnewline
\multicolumn{1}{l}{\ h \ m \ \ \ s} & \multicolumn{1}{l}{\ \ \ $\degree$ \ \ \ $\arcmin$ \ \ \ $\arcsec$} &   (mas)   &   (mas)  & (jd) & & & & \tabularnewline
\hline
 16 59 11.6508 & -22 00 44.855 &  17 &  12 & 2454147.78241319 &   16.0 &    C &      BC & 874 \tabularnewline
 16 59 11.6845 & -22 00 44.932 &  17 &  12 & 2454147.78332384 &   15.8 &    C &      BC & 874 \tabularnewline
 16 59 11.7181 & -22 00 44.978 &  17 &  12 & 2454147.78422477 &   16.0 &    C &      BC & 874 \tabularnewline
 16 59 11.7818 & -22 00 45.143 &  17 &  12 & 2454147.78602662 &   15.9 &    C &      BC & 874 \tabularnewline
 16 59 11.8188 & -22 00 45.232 &  17 &  12 & 2454147.78693750 &   16.0 &    C &      BC & 874 \tabularnewline
 17 17 11.0344 & -22 47 19.415 &  30 &  24 & 2454205.63885463 &   16.1 &        U &      BC & 874 \tabularnewline
 17 17 11.0270 & -22 47 19.381 &  30 &  24 & 2454205.63959167 &   16.1 &        U &      BC & 874 \tabularnewline
 17 17 11.0258 & -22 47 19.366 &  30 &  24 & 2454205.64031875 &   16.1 &        U &      BC & 874 \tabularnewline
 17 17 11.0192 & -22 47 19.417 &  30 &  24 & 2454205.64104583 &   16.1 &        U &      BC & 874 \tabularnewline
\hline
\end{tabular}
\par\end{centering}
This sample corresponds to 9 observations of Himalia from February 16, 2007 and April 15, 2007. Tables contain the topocentric ICRS coordinates of the irregular satellites, the position error estimated from the dispersion of the ephemeris offsets of the night of observation, the UTC time of the frame's mid-exposure in Julian date, the estimated magnitude, the filter used, the telescope origin and correspondent IAU code. The filters may be U, B, V, R or I following the Johnson system; C stands for clear (no filter used), resulting in a broader R band magnitude, RE for the broad-band R filter ESO\#844 with $\lambda_{c} = 651.725$ nm and $\Delta\lambda = 162.184 $ nm (full width at half maximum) and "un" for unknown filter. E, OH, PE, BC and Z stand for the ESO, OHP, Perkin-Elmer, Bollen \& Chivens and Zeiss telescopes, respectively.
\end{table*}

\begin{table}
\caption{\label{Tab: comparison-horizons} Comparison of positions obtained with \citealp{Jacobson2012}.}
\begin{centering}
\begin{tabular}{lccccc}
\hline  \hline
 & \multicolumn{4}{c}{Number of Positions}   &    \tabularnewline
Satellite  & OPD  & OHP & ESO & Total  & Jacobson \tabularnewline
\hline
Himalia & 854 & 357 & 23 & 1234 & 1757 \tabularnewline
Elara & 403 & 187 & 46 & 636 & 1115 \tabularnewline
Lysithea & 60 & 84 & 90 & 234 & 431 \tabularnewline
Leda & 6 & 48 & 44 & 98 & 178 \tabularnewline
\hdashline
Pasiphae & 295 & 248 & 66 & 609 & 1629 \tabularnewline
Callirrhoe & 9 & -  &  16 & 25 & 95 \tabularnewline
Megaclite & - & -  &  10 & 10 & 50  \tabularnewline
\hdashline
Ananke & 52 & 141 & 57 & 250 & 600 \tabularnewline
Praxidike & - & -  &   2 & 2 & 59 \tabularnewline
\hdashline
Carme & 90 & 204 & 37 & 331 & 973 \tabularnewline
Sinope & 41 & 169 & 11 & 221 & 854 \tabularnewline
Themisto & - & - & 16 & 16 & 55 \tabularnewline
\hline
Phoebe & 1239 & 516 & 32 & 1787 & 3479 \tabularnewline
\hdashline
Siarnaq & - & 20 & 56 & 76 & 239 \tabularnewline
Paaliaq & - & - & 11 & 11 & 82 \tabularnewline
\hdashline
Albiorix & - & - & 46 & 46 & 137 \tabularnewline
\hline
Sycorax & - & - & 35 & 35 & 237 \tabularnewline
\hline
Nereid & 803 & - & 99 & 902 & 716 \tabularnewline
\hline
\end{tabular}
\par\end{centering}
Comparison between the number of positions obtained in our work with the number used in the numerical integration of orbits by the JPL as published by \citealp{Jacobson2012}.
\end{table}

\section{Comparison with ephemeris} \label{Sec: comparison}

\begin{figure*}
\begin{centering}
\subfigure[Right Ascension]{\includegraphics[scale=0.5]{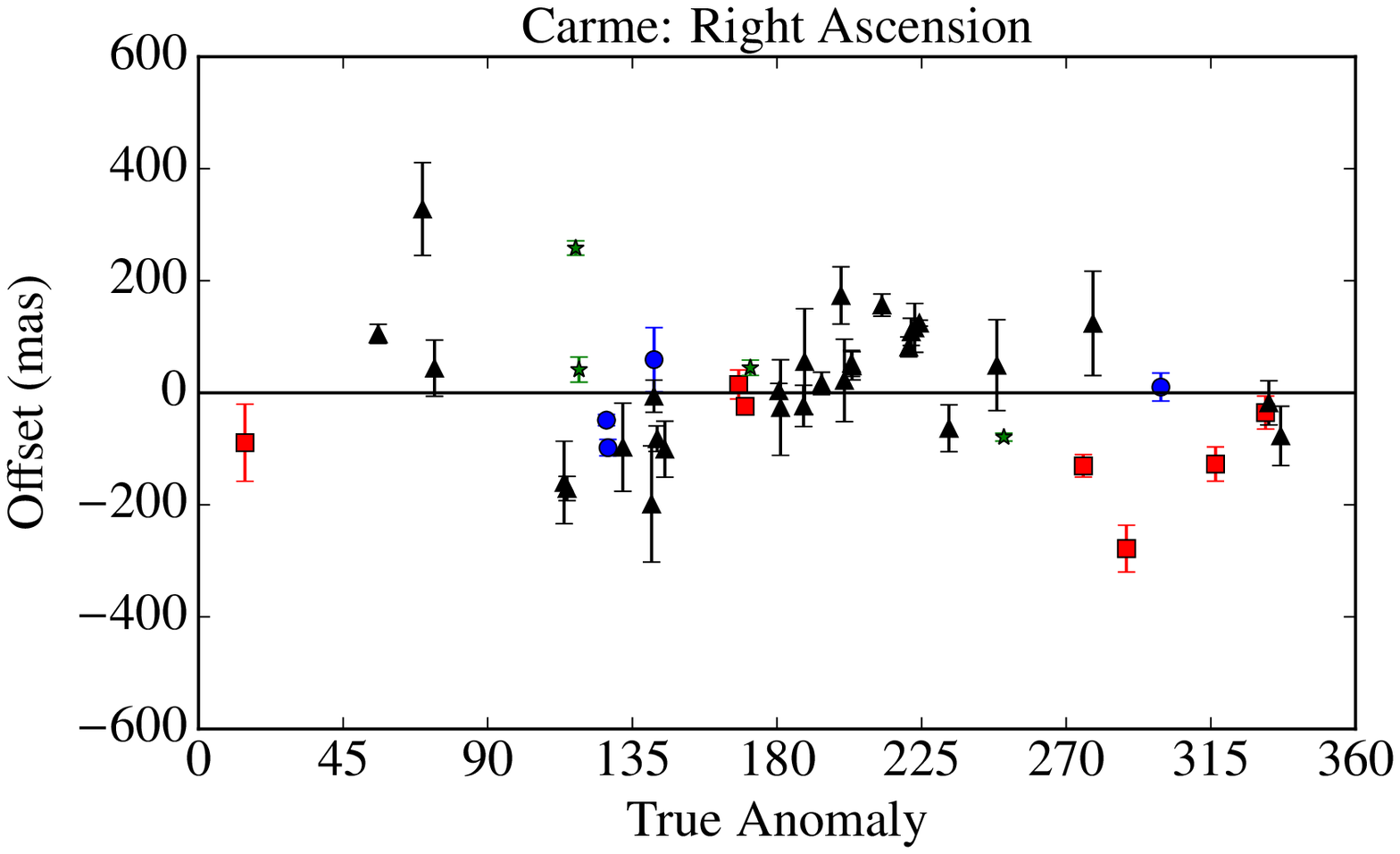}\label{Fig: carme_alfa}}
\subfigure[Declination]{\includegraphics[scale=0.5]{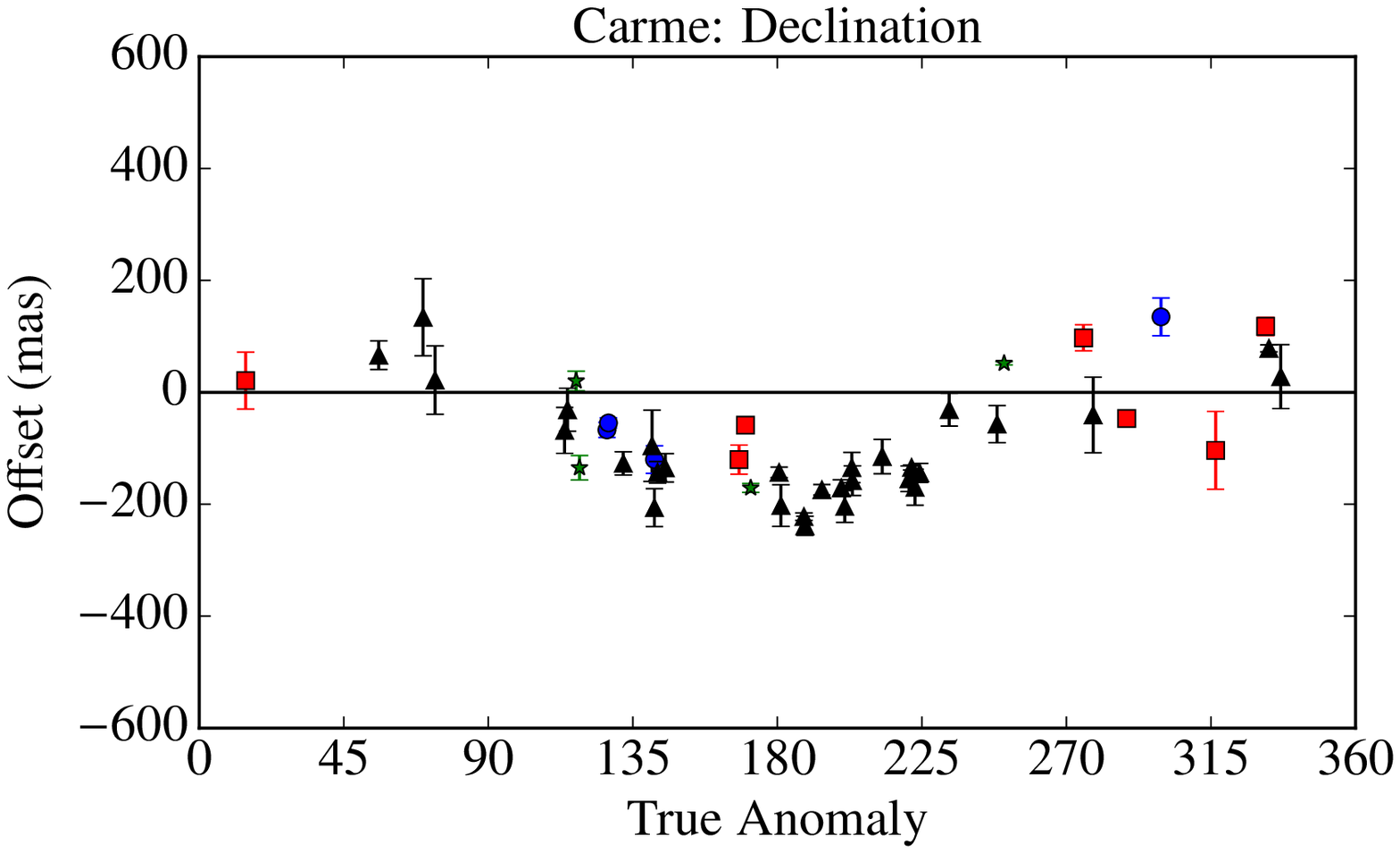}\label{Fig: carme_delta}}
\caption{Mean ephemeris offsets and dispersions (1 sigma error bars) in the coordinates of Carme taken night by night by true anomaly for each telescope. The red square is for the observations with the Perkin-Elmer telescope from OPD, the blue circle for Boller \& Chivens, the magenta triangle down for Zeiss, the black triangle up for OHP and the green star for ESO.}
\label{Fig: carme_anom}
\end{centering}
\end{figure*}

Intending to see the potential of our results to improve the orbit of the irregular satellites observed, we analyzed the offsets of our positions with regard to the ephemeris mentioned in Sect. \ref{Sec: reduction}. Taking Carme as example, we plot the mean ephemeris offsets for each night in Fig. \ref{Fig: carme_anom} and their dispersions  (one sigma error bars) as a function of the true anomaly in right ascension (\ref{Fig: carme_alfa}) and declination (\ref{Fig: carme_delta}). Figure \ref{Fig: carme_delta} clearly shows a systematic error in declination. When Carme is close to its apojove (true anomaly = 180\degree), its offsets are more likely to be more negative than those close to its perijove (true anomaly = 0\degree). The offsets obtained from observations by four telescopes using different cameras and filters are in good agreement, meaning that there is an error in the ephemeris of Carme, most probably due to an error in its orbital inclination.

\begin{figure*}
\begin{centering}
\subfigure[Right Ascension]{\includegraphics[scale=0.5]{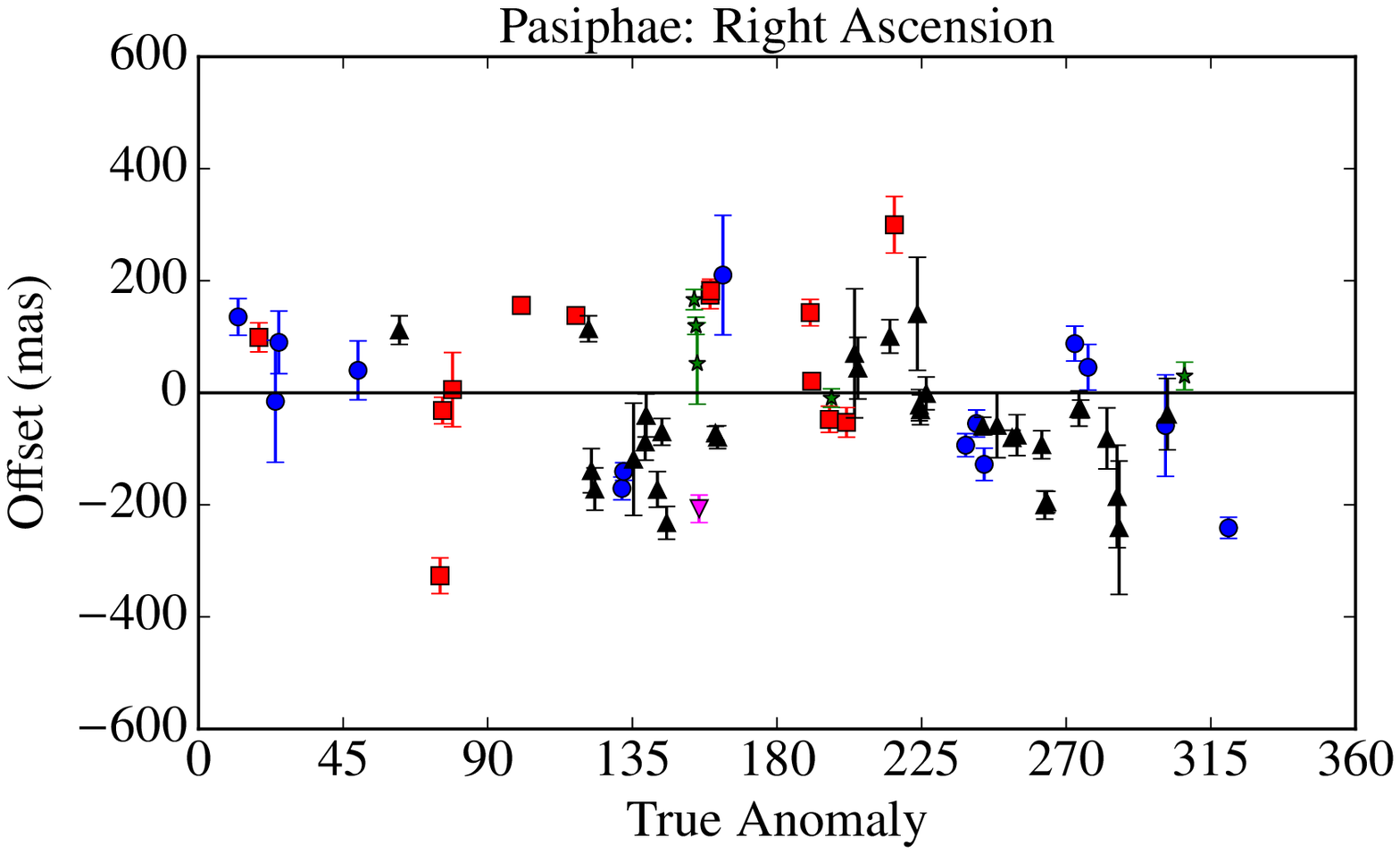}\label{Fig: pasiphae_alfa}}
\subfigure[Declination]{\includegraphics[scale=0.5]{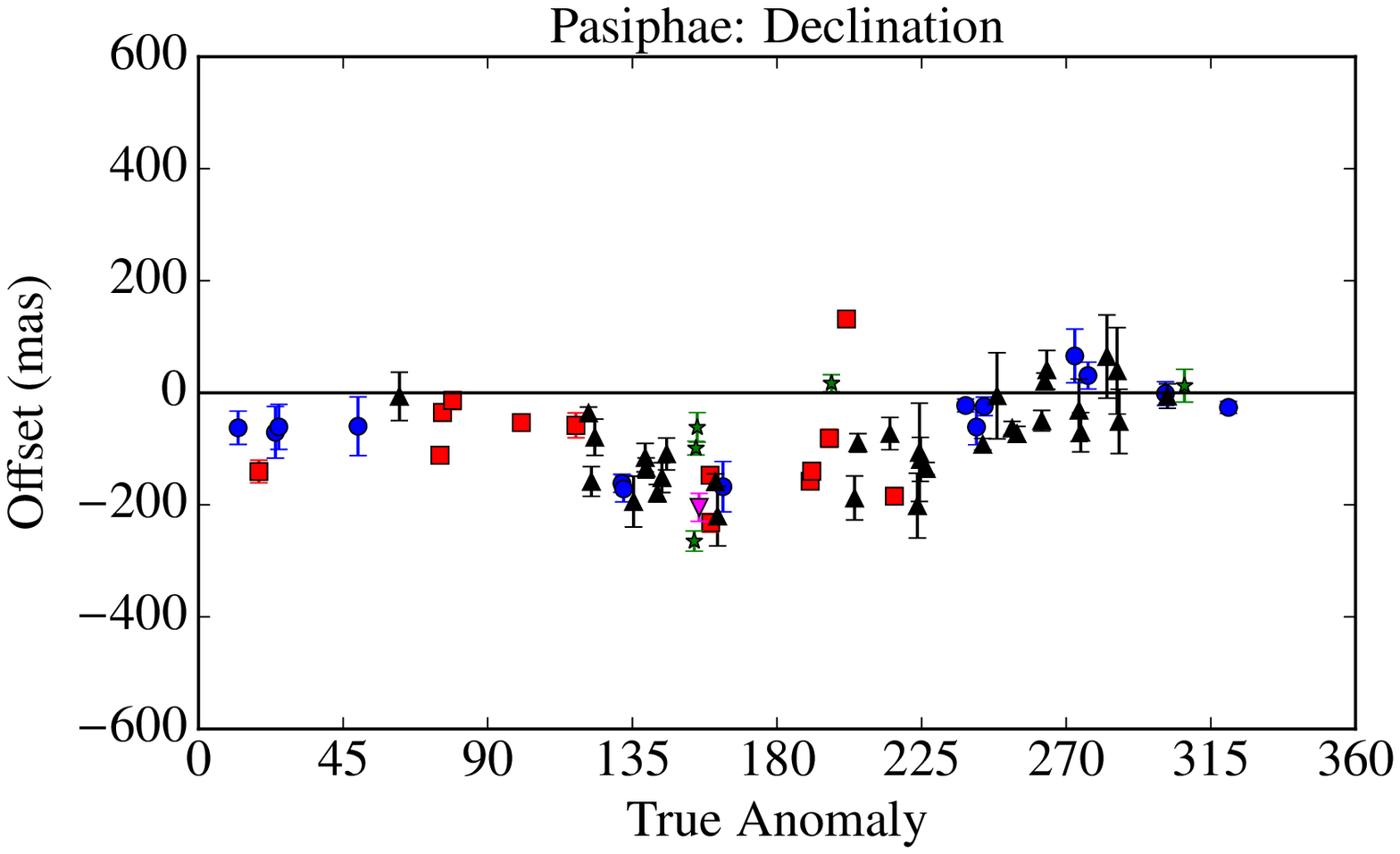}\label{Fig: pasiphae_delta}}
\caption{Same as in Fig \ref{Fig: carme_anom} for Pasiphae.}
\label{Fig: pasiphae_anom}
\end{centering}
\end{figure*}

This pattern in declination was also seen for other satellites like Pasiphae (Fig: \ref{Fig: pasiphae_anom}) and Ananke (plots for other satellites with significant number of observations can be seen in the online material). For some satellites, the orbital coverage is not enough to clearly indicate the presence of systematic errors in specific orbital elements. However, after comparing the internal position mean errors of the reductions (Table \ref{Tab: stars-errors}) with the external position errors estimated from the dispersion of the ephemeris offsets (Tables \ref{Tab: Reductions-160} to \ref{Tab: Reductions-eso}), we see position error values that are much higher than expected from the mean errors. This means that besides some expected astrometric errors, significant ephemeris errors must also be present.

\section{Conclusions} \label{Sec: conclusions}

We managed a large database with FITS images acquired by five telescopes in three sites between 1992 and 2014. From that, we identified 8466 observations of irregular satellites, from which we managed to obtain 6523 suitable astrometric positions, giving a total of 3666 positions for 12 satellites of Jupiter, 1920 positions for 4 satellites of Saturn, 35 positions for Sycorax (Uranus) and 902 positions for Nereid (Neptune).

The positions of all the objects were determined using the PRAIA package. The package was suited to coping with the huge number of observations and with the task of identifying the satellites within the database. PRAIA tasks were also useful for dealing with the missing or incorrect coordinate and time stamps present mostly in the old observations.

The UCAC4 was used as the reference frame. Based on the comparisons with ephemeris, we estimate that the position errors are about 60 mas to 80 mas depending on the satellite brightness. For some satellites the number of positions obtained in this work is comparable to the number used in the numerical integration of orbits by the JPL \citep{Jacobson2012} (see Table \ref{Tab: comparison-horizons}). For instance, the number of new positions for Himalia is about 70\% of the number used in the numerical integation of orbits by JPL. Systematic errors in the ephemeris were found for at least some satellites (Ananke, Carme, Elara and Pasiphae). In the case of Carme, we showed an error in the orbital inclination (see Fig. \ref{Fig: carme_anom}). 

The positions derived in this work can be used in new orbital numerical integrations, generating more precise ephemerides. Stellar occultations by irregular satellites could then be predicted better. Based on this work, our group has already computed occultation predictions for the eight major irregular satellites of Jupiter. These predictions will be published in a forthcoming paper.

\begin{acknowledgements}

ARGJ acknowledges the financial support of CAPES. M. A. is grateful to the CNPq (Grants 473002/2013-2 and 308721/2011-0) and FAPERJ (Grant E-26/111.488/2013). RVM acknowledges the following grants: CNPq-306885/2013, Capes/Cofecub-2506/2015, Faperj/PAPDRJ-45/2013. J-E.A. is grateful to the "Programme National de Planétologie" of INSU-CNRS-CNES for its financial support. J.I.B. Camargo acknowledges CNPq for a PQ2 fellowship (process number 308489/2013-6). F.B.R. acknowledges PAPDRJ-FAPERJ/CAPES E-43/2013 number 144997, E-26/101.375/2014. B.E.M. is grateful for the financial support of CAPES.

\end{acknowledgements}

\bibliographystyle{aa}
\nocite{*}
\bibliography{references.bib}

\Online

\begin{appendix}

\onecolumn

\section{Ephemeris offsets as a function of true anomaly for all observed irregular satellites} \label{anexo: offsets}

The distribution of ephemeris offsets along the orbit of the satellites are shown below. The red square is for the observations with the Perkin-Elmer telescope from OPD, the blue circle for Boller \& Chivens, the magenta triangle down for Zeiss, the black triangle up for OHP and the green star for ESO. For Carme and Pasiphae see Figs. \ref{Fig: carme_anom} and \ref{Fig: pasiphae_anom} in Section \ref{Sec: comparison}.

\begin{figure}[h!]
\begin{centering}
\subfigure[Right Ascension]{\includegraphics[scale=0.5]{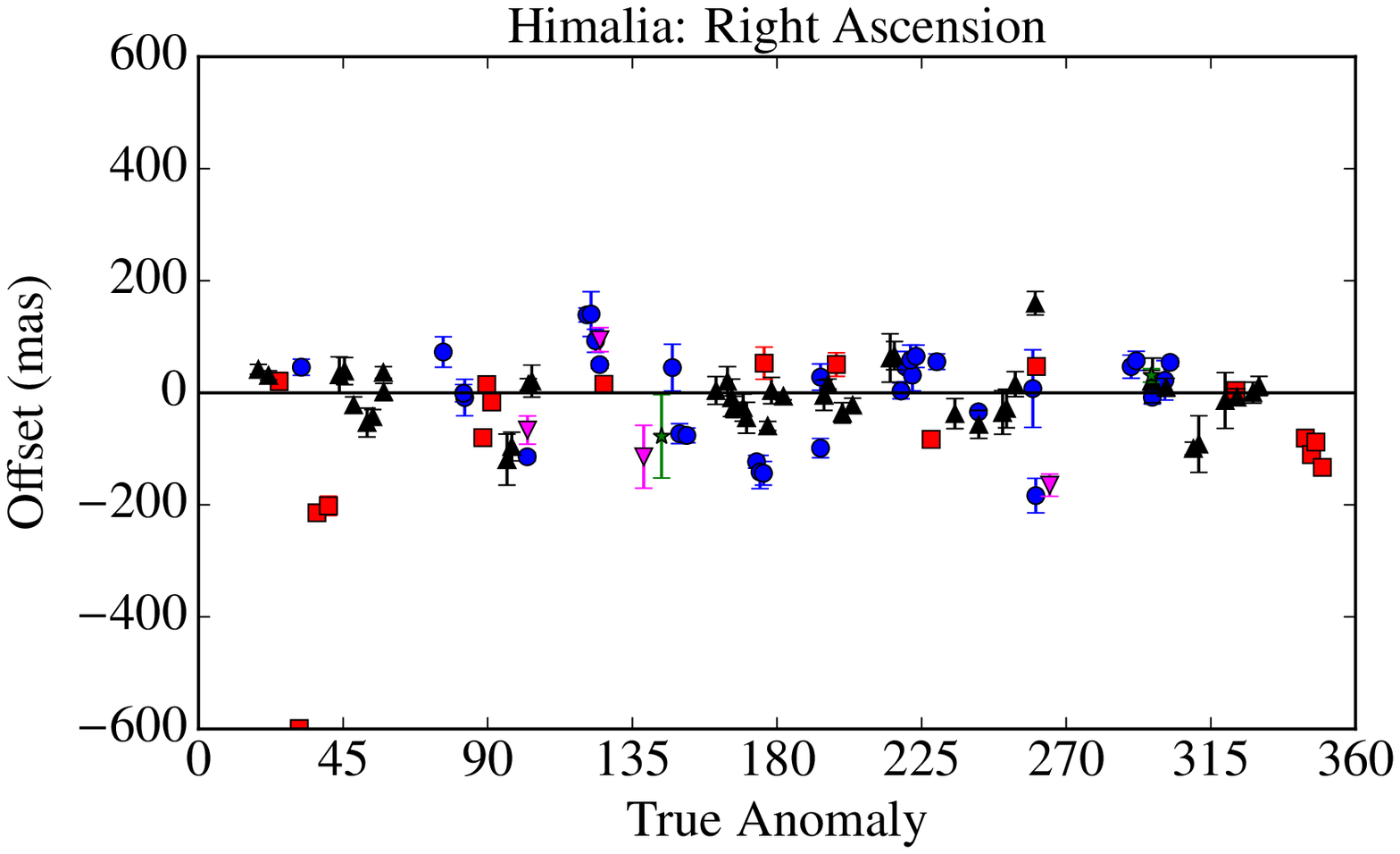}\label{Fig: an_himalia_alfa}}
\subfigure[Declination]{\includegraphics[scale=0.5]{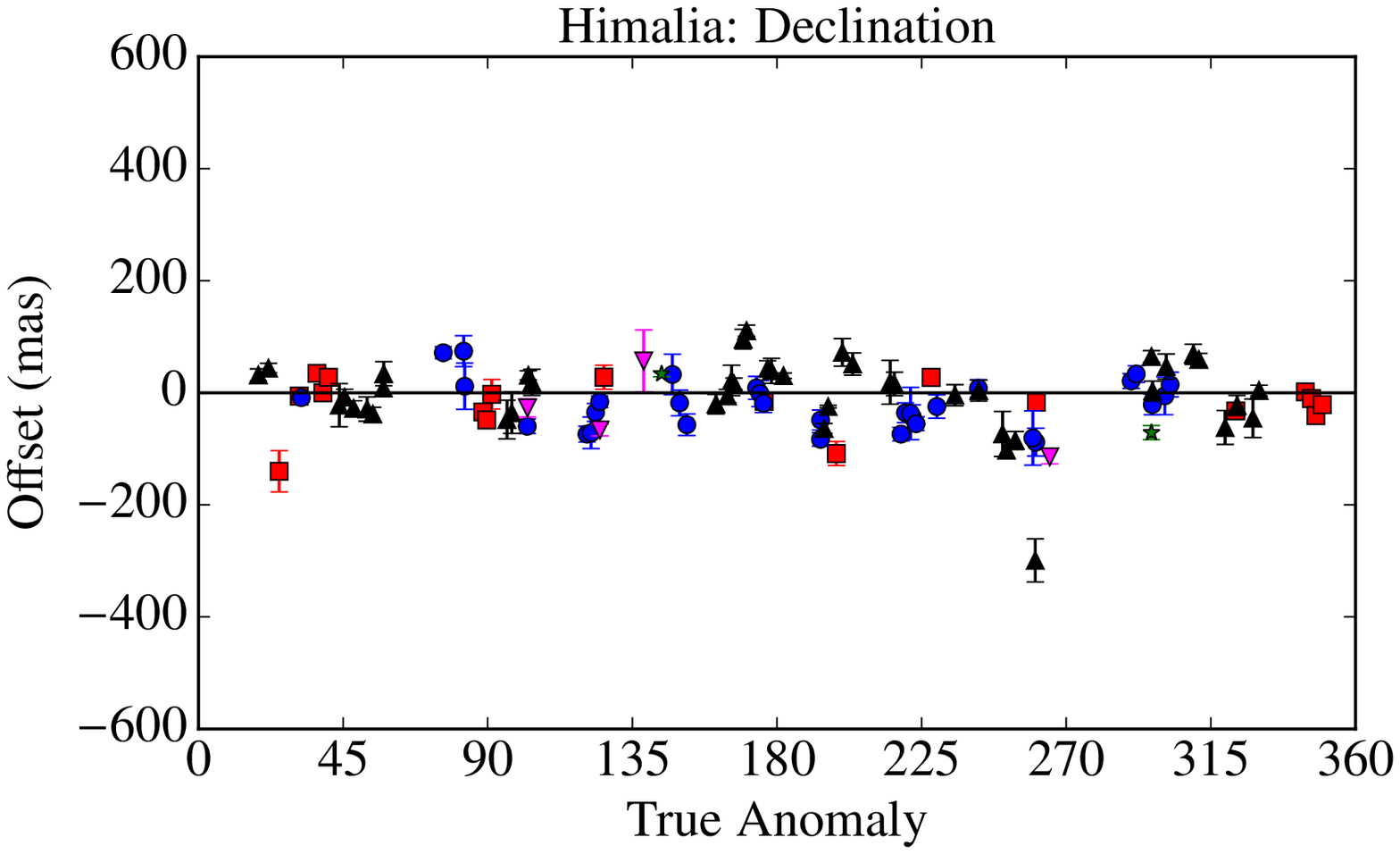}\label{Fig: an_himalia_delta}}
\caption{Mean ephemeris offset and dispersion (1 sigma error bars) in the coordinates of Himalia taken night by night as a function of true anomaly.}
\label{Fig: an_himalia_anom}
\end{centering}
\end{figure}

\begin{figure}[h!]
\begin{centering}
\subfigure[Right Ascension]{\includegraphics[scale=0.5]{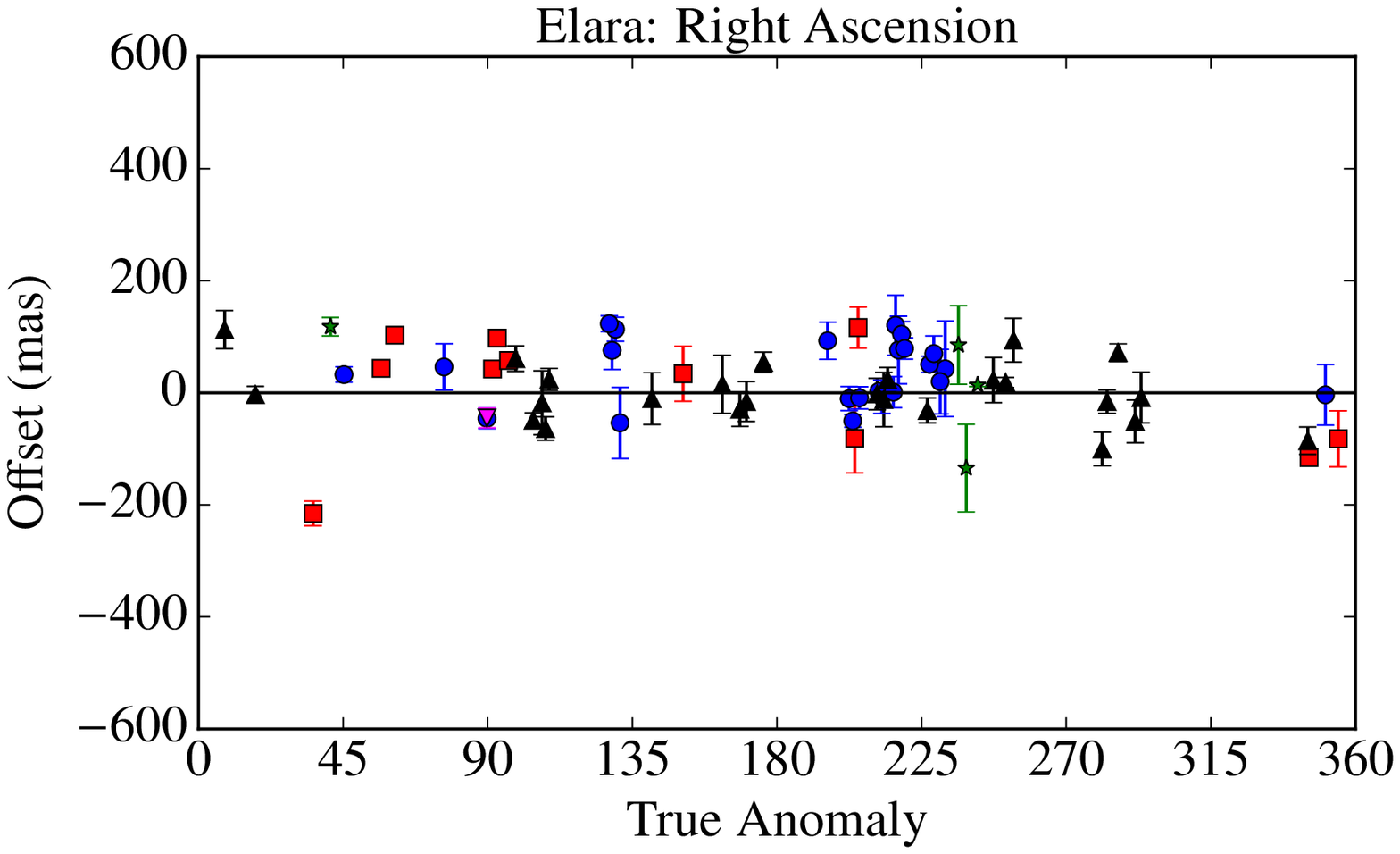}\label{Fig: an_elara_alfa}}
\subfigure[Declination]{\includegraphics[scale=0.5]{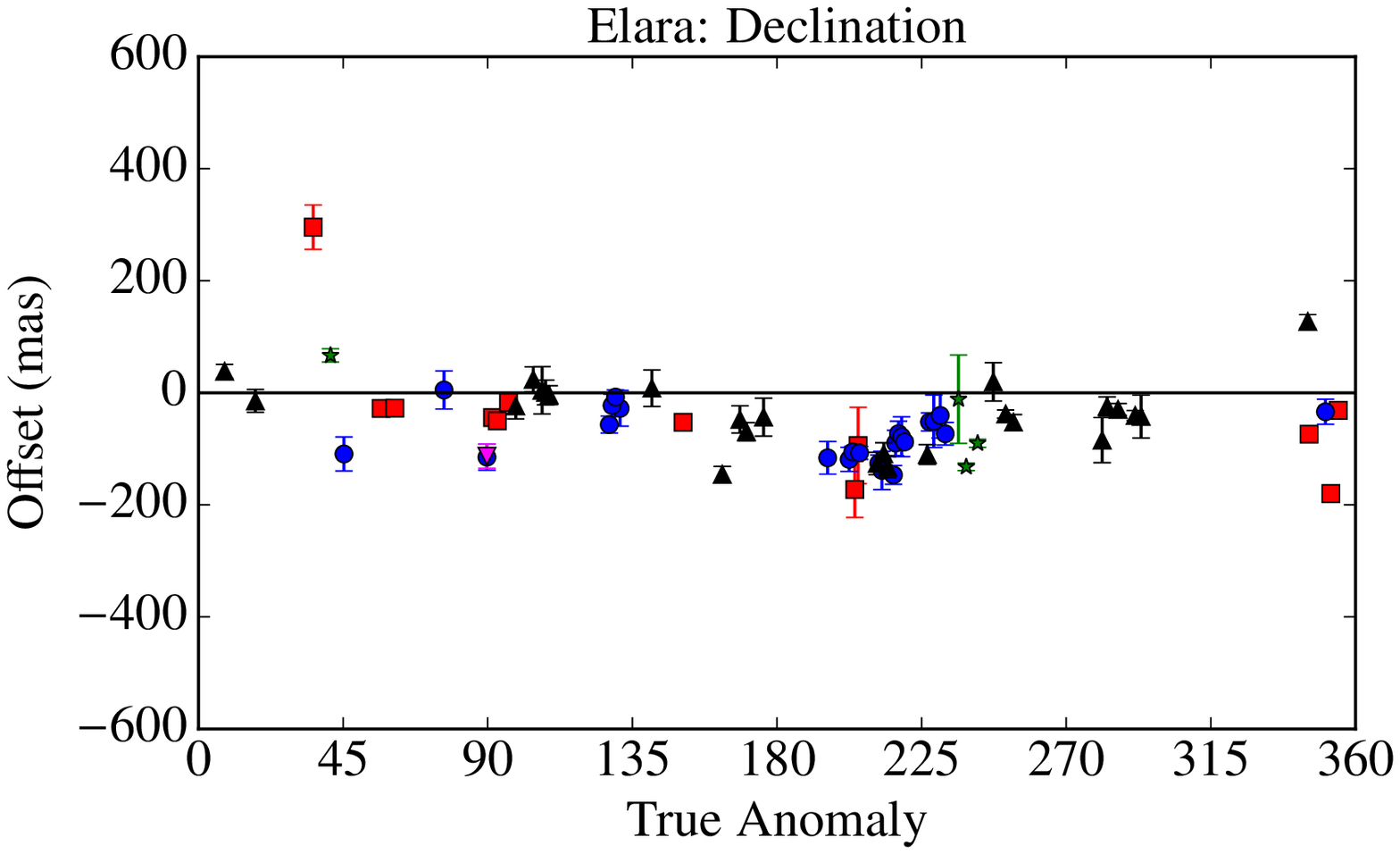}\label{Fig: an_elara_delta}}
\caption{Same as in Fig \ref{Fig: an_himalia_anom} for Elara.}
\label{Fig: an_elara_anom}
\end{centering}
\end{figure}

\begin{figure}[h!]
\begin{centering}
\subfigure[Right Ascension]{\includegraphics[scale=0.5]{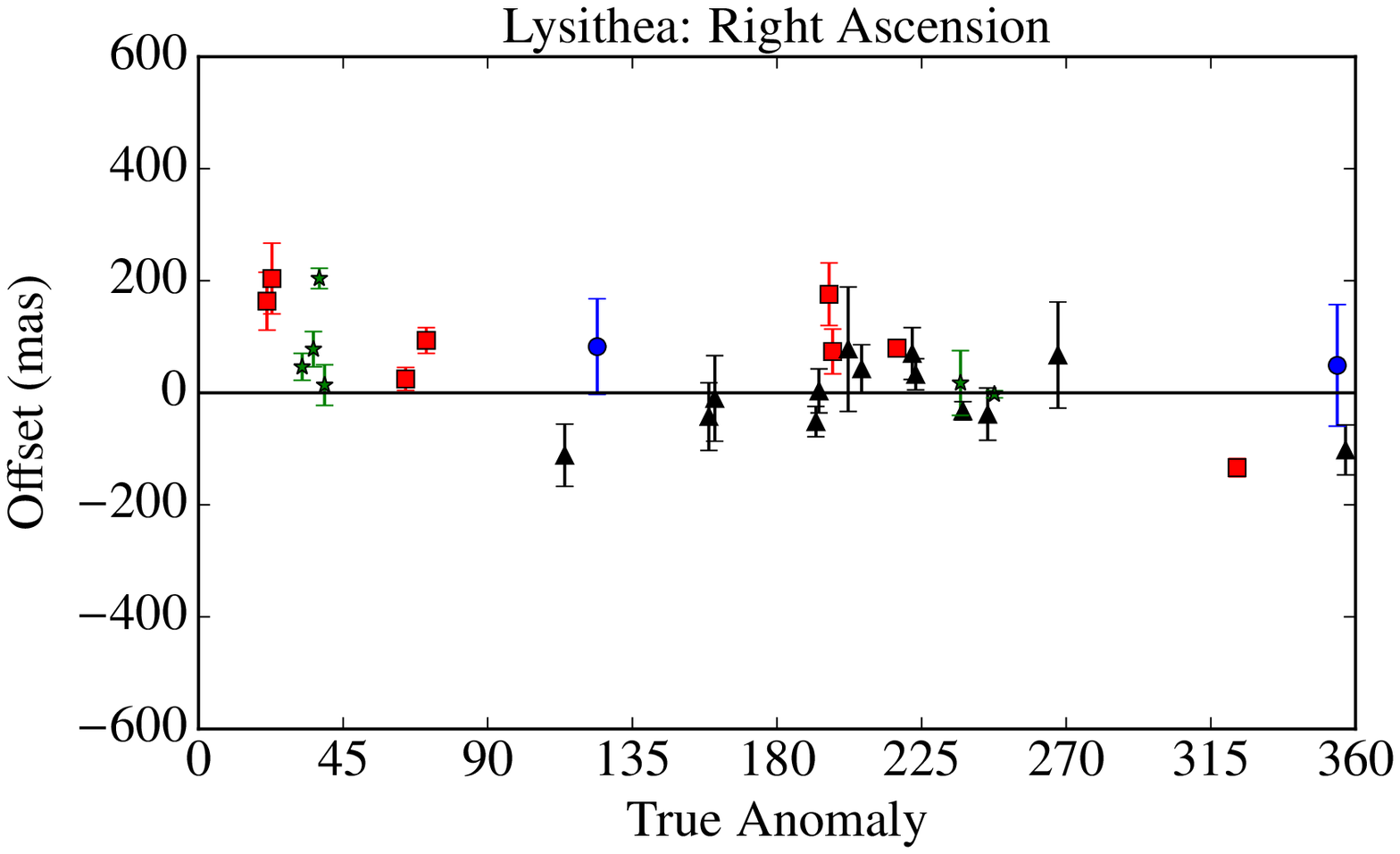}\label{Fig: an_lysithea_alfa}}
\subfigure[Declination]{\includegraphics[scale=0.5]{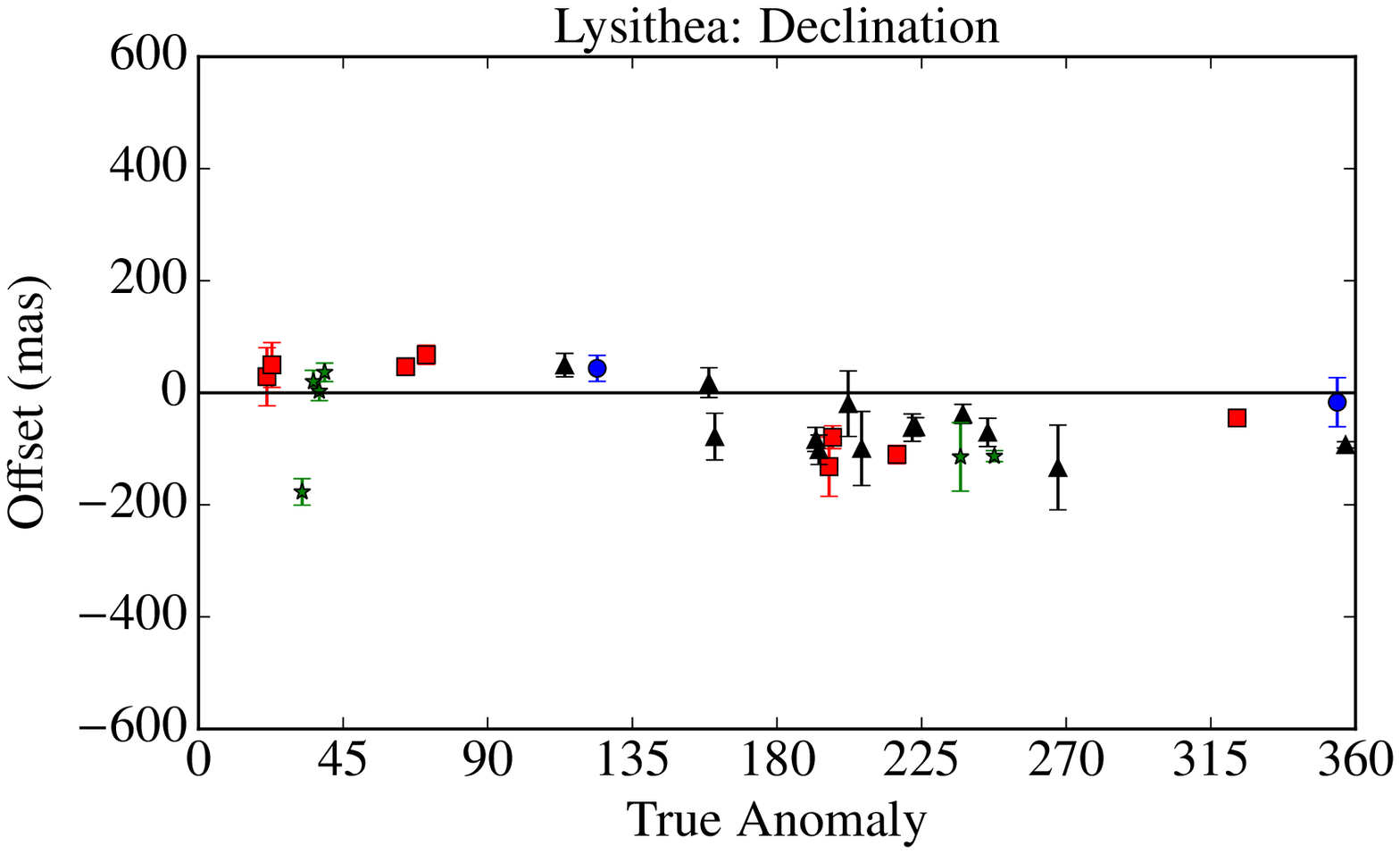}\label{Fig: an_lysithea_delta}}
\caption{Same as in Fig \ref{Fig: an_himalia_anom} for Lysithea.}
\label{Fig: an_lysithea_anom}
\end{centering}
\end{figure}

\begin{figure}[h!]
\begin{centering}
\subfigure[Right Ascension]{\includegraphics[scale=0.5]{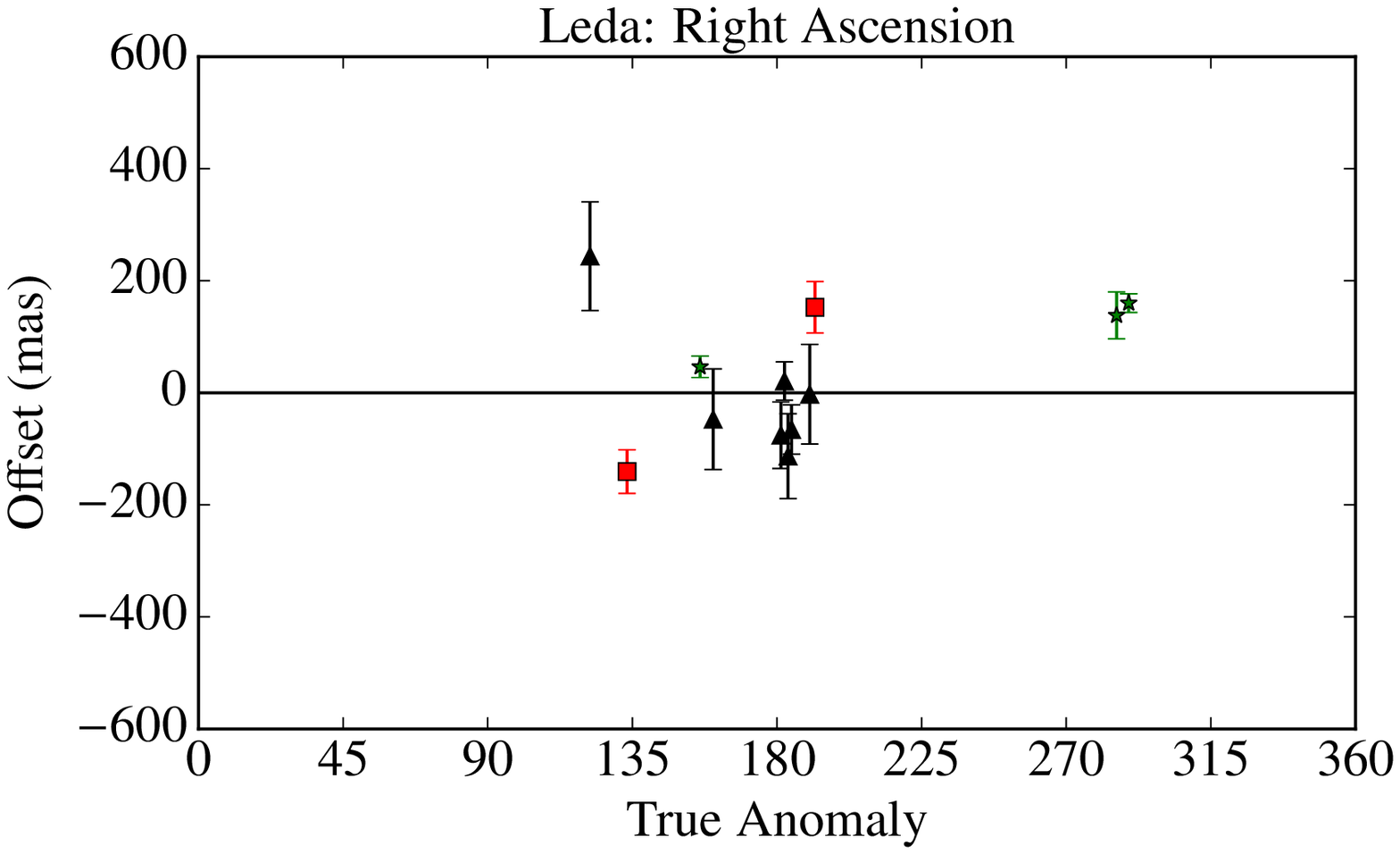}\label{Fig: an_leda_alfa}}
\subfigure[Declination]{\includegraphics[scale=0.5]{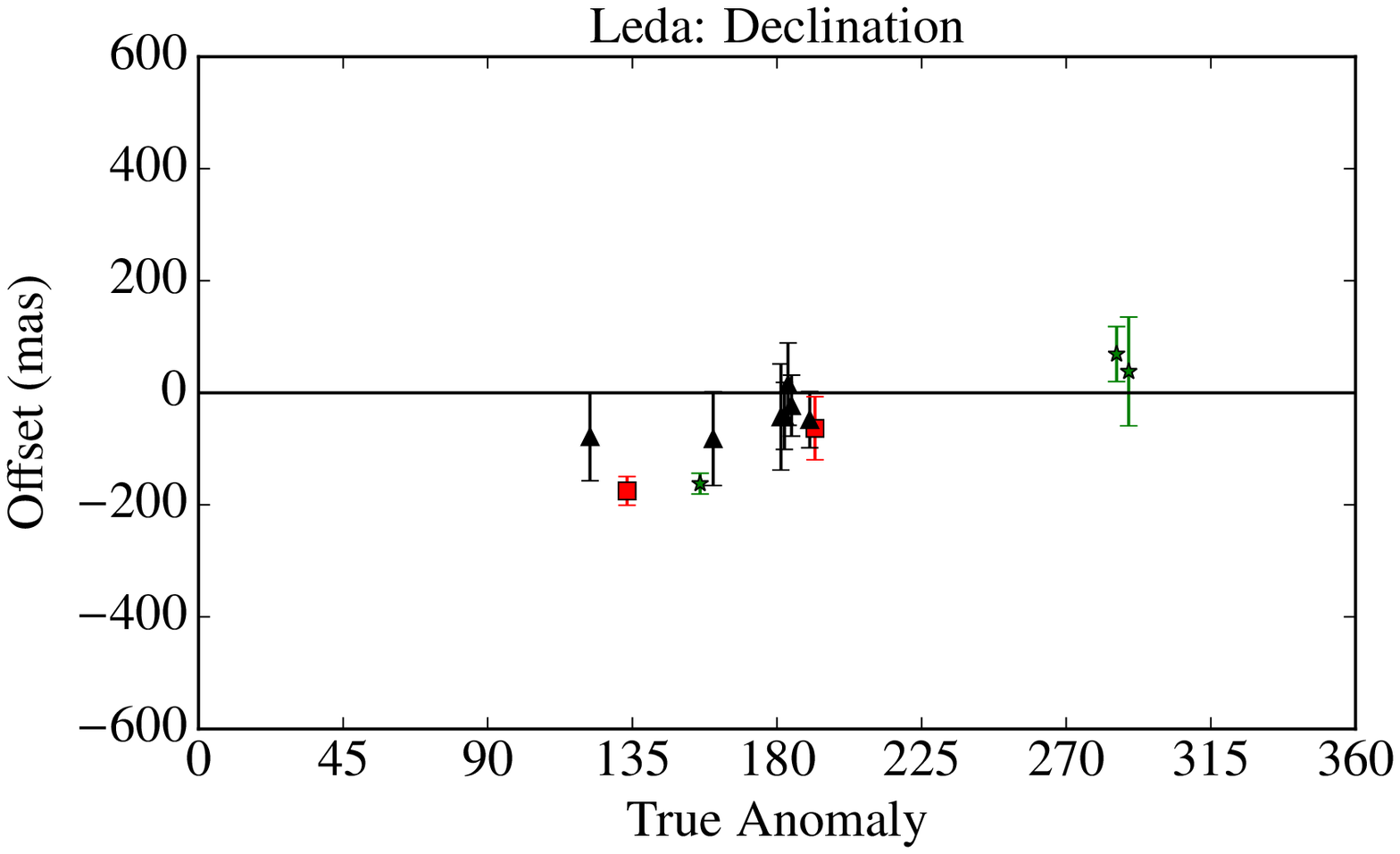}\label{Fig: an_leda_delta}}
\caption{Same as in Fig \ref{Fig: an_himalia_anom} for Leda.}
\label{Fig: an_leda_anom}
\end{centering}
\end{figure}

\begin{figure}[h!]
\begin{centering}
\subfigure[Right Ascension]{\includegraphics[scale=0.5]{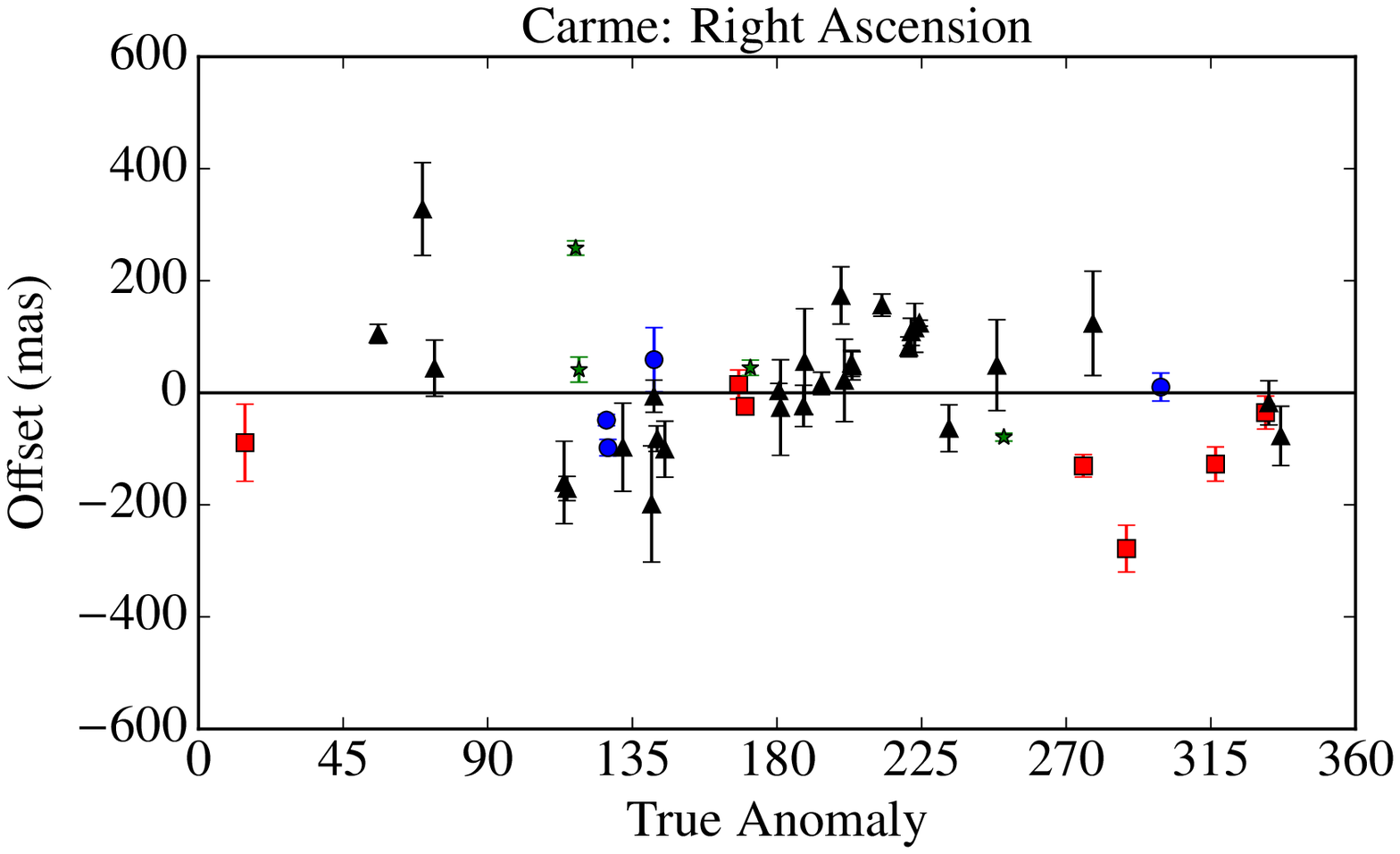}\label{Fig: an_carme_alfa}}
\subfigure[Declination]{\includegraphics[scale=0.5]{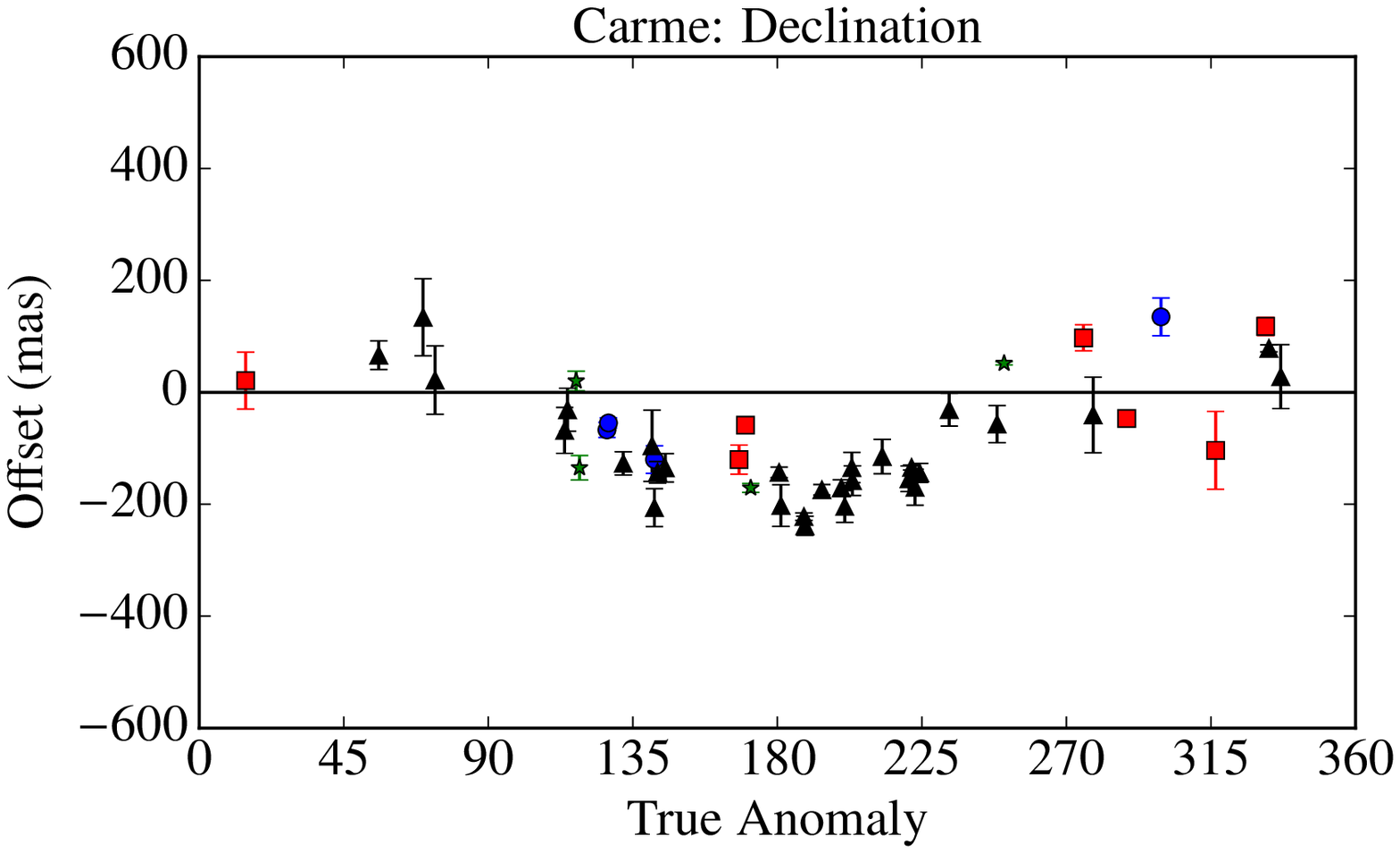}\label{Fig: an_carme_delta}}
\caption{Same as in Fig \ref{Fig: an_himalia_anom} for Carme.}
\label{Fig: an_carme_anom}
\end{centering}
\end{figure}

\begin{figure}[h!]
\begin{centering}
\subfigure[Right Ascension]{\includegraphics[scale=0.5]{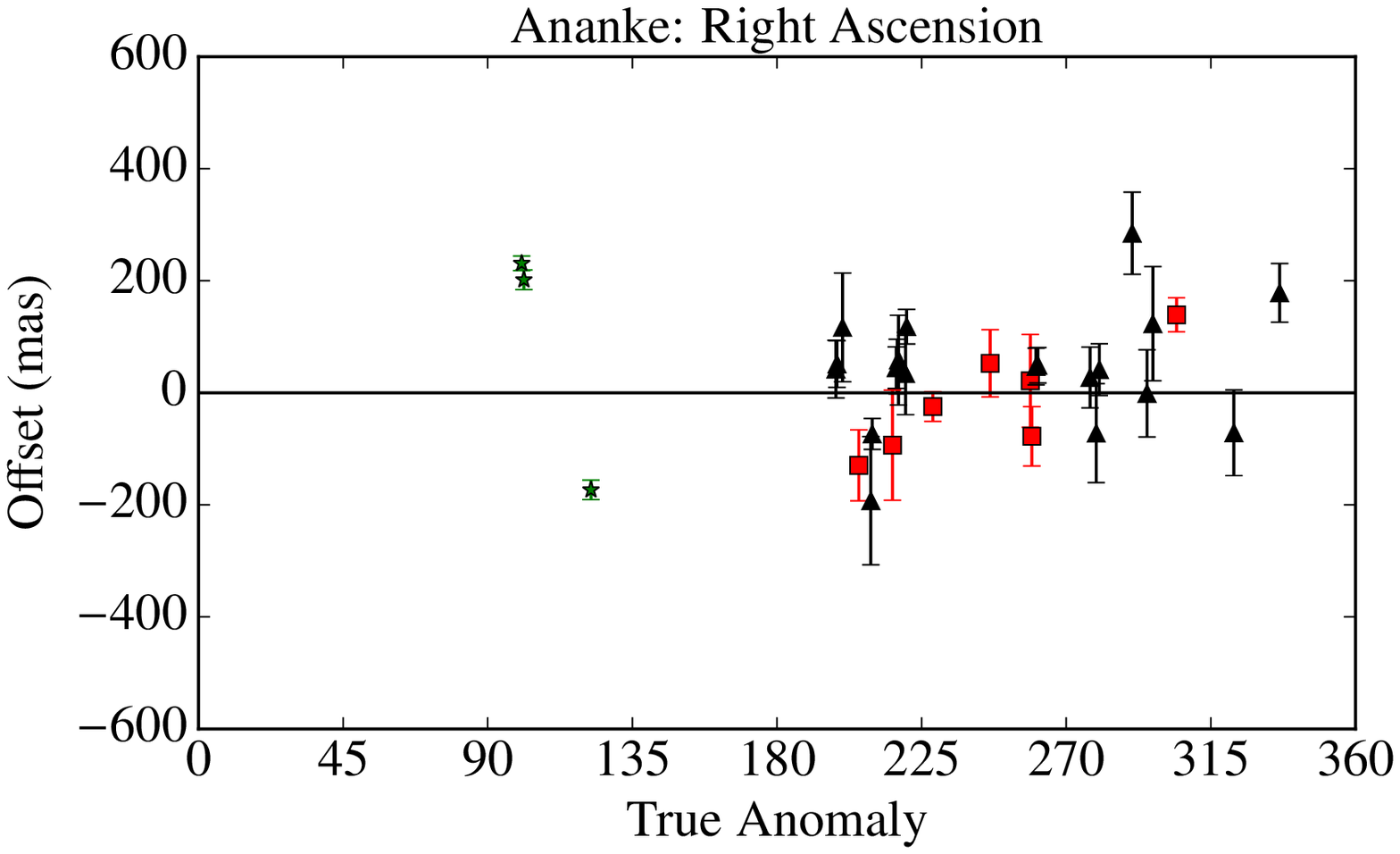}\label{Fig: an_ananke_alfa}}
\subfigure[Declination]{\includegraphics[scale=0.5]{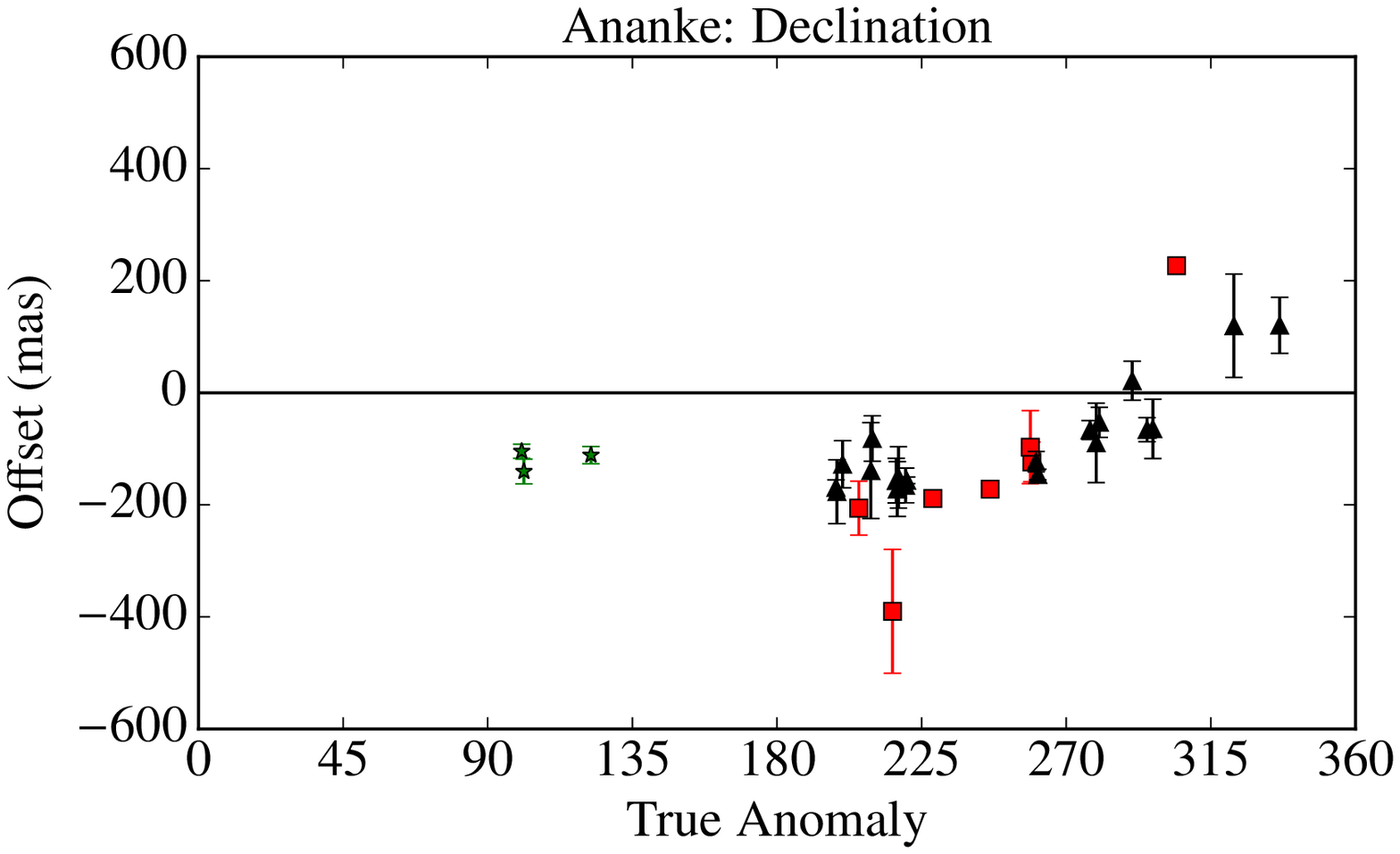}\label{Fig: an_ananke_delta}}
\caption{Same as in Fig \ref{Fig: an_himalia_anom} for Ananke.}
\label{Fig: an_ananke_anom}
\end{centering}
\end{figure}

\begin{figure}[h!]
\begin{centering}
\subfigure[Right Ascension]{\includegraphics[scale=0.5]{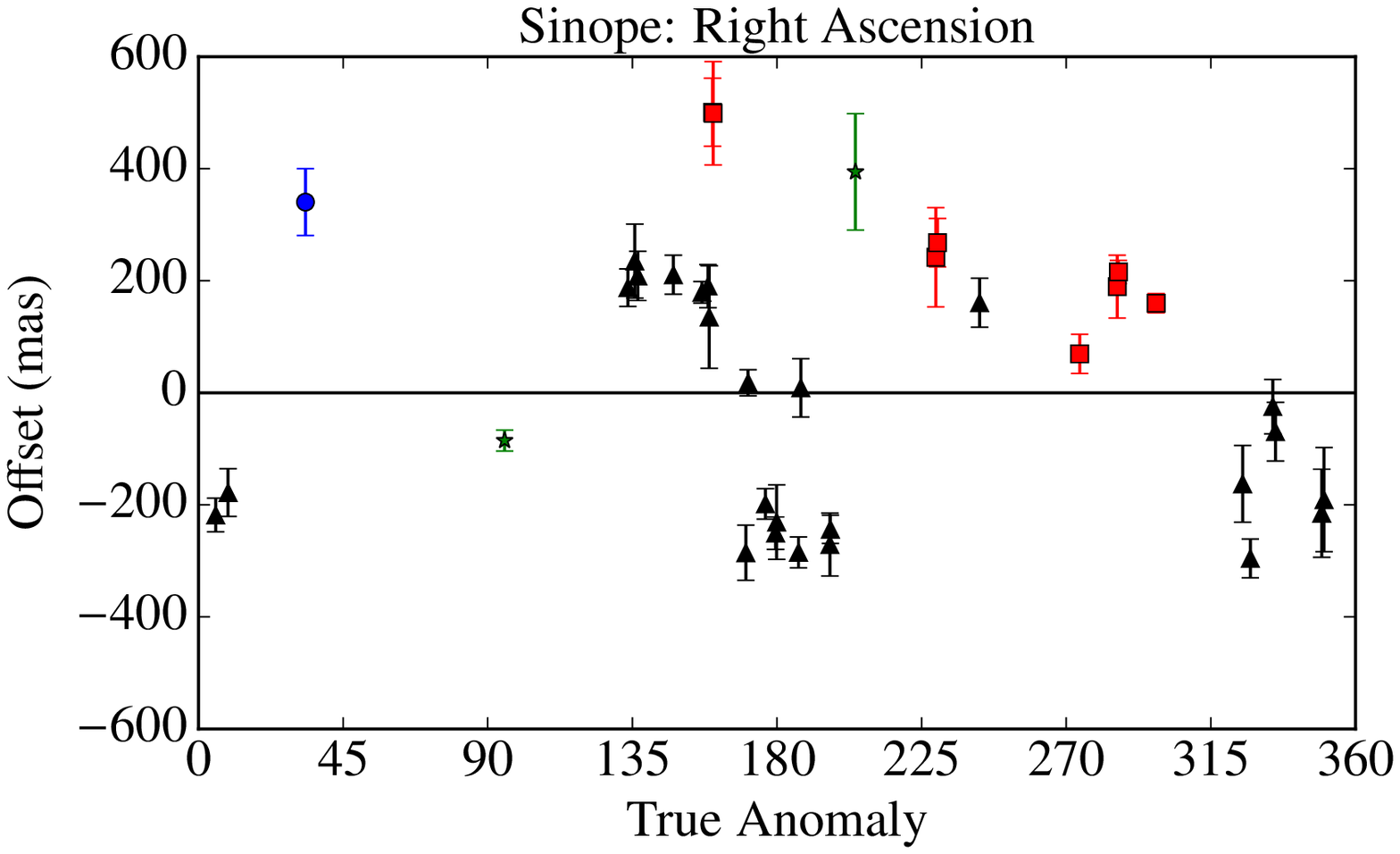}\label{Fig: an_sinope_alfa}}
\subfigure[Declination]{\includegraphics[scale=0.5]{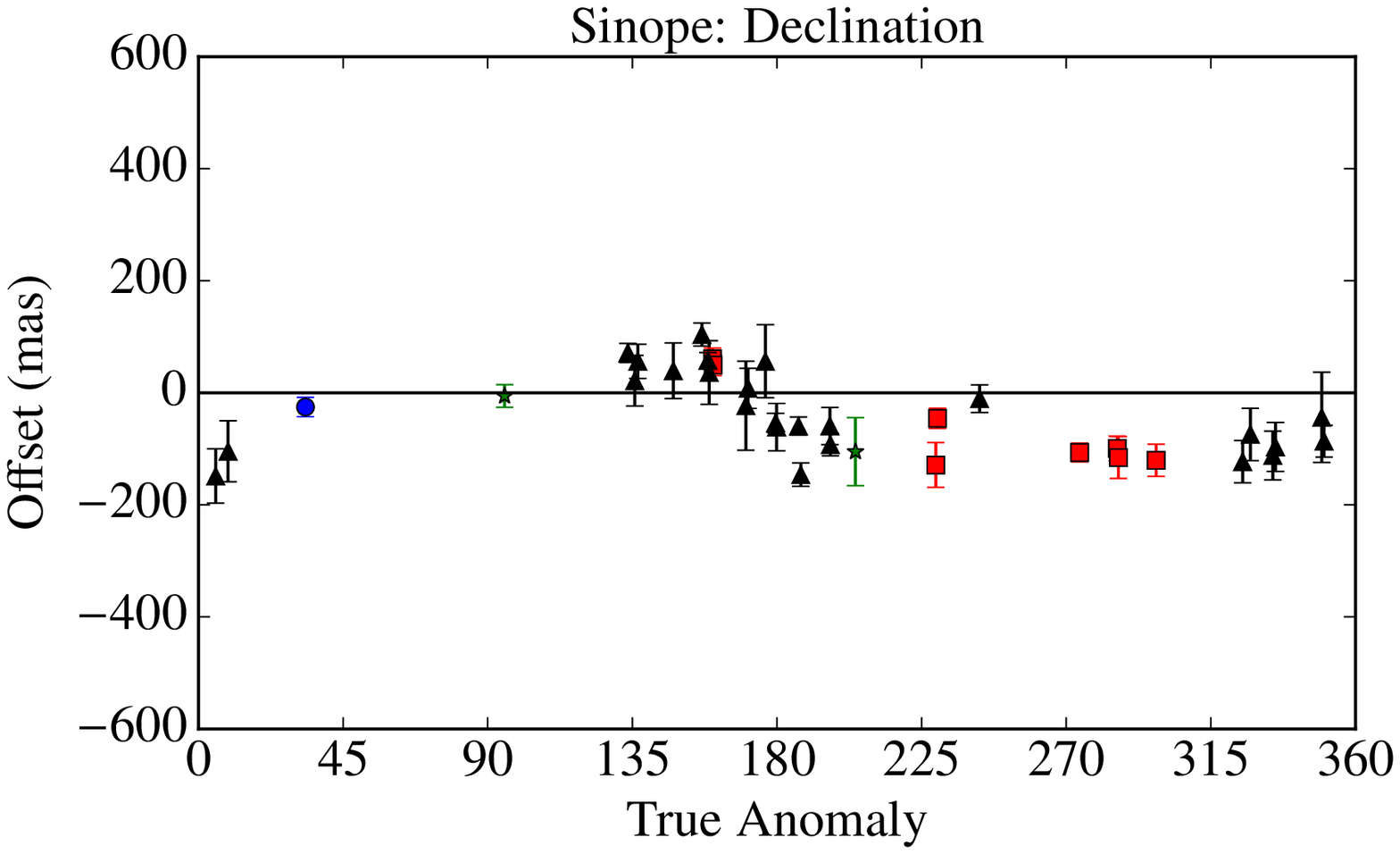}\label{Fig: an_sinope_delta}}
\caption{Same as in Fig \ref{Fig: an_himalia_anom} for Sinope.}
\label{Fig: an_sinope_anom}
\end{centering}
\end{figure}

\begin{figure}[h!]
\begin{centering}
\subfigure[Right Ascension]{\includegraphics[scale=0.5]{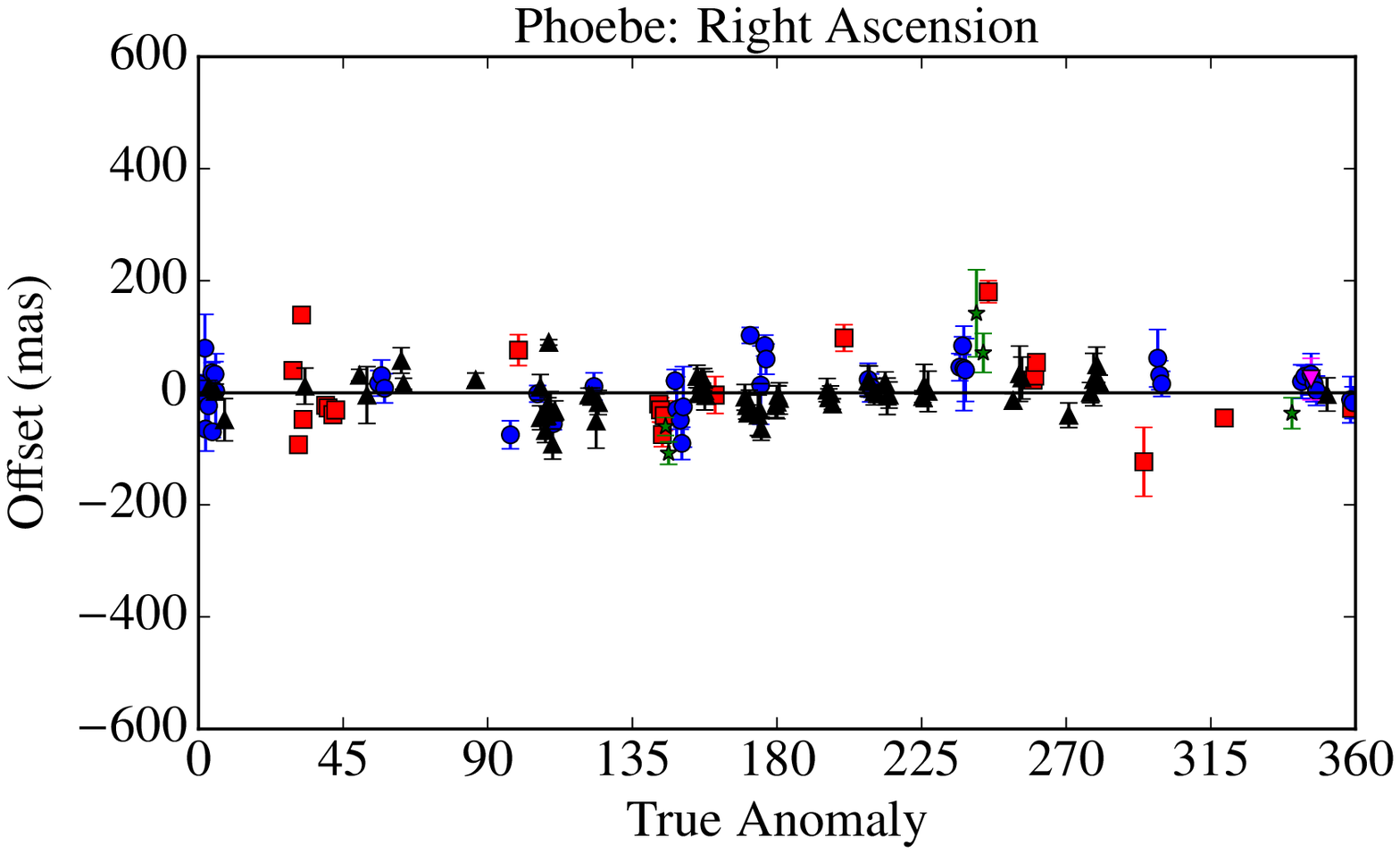}\label{Fig: an_phoebe_alfa}}
\subfigure[Declination]{\includegraphics[scale=0.5]{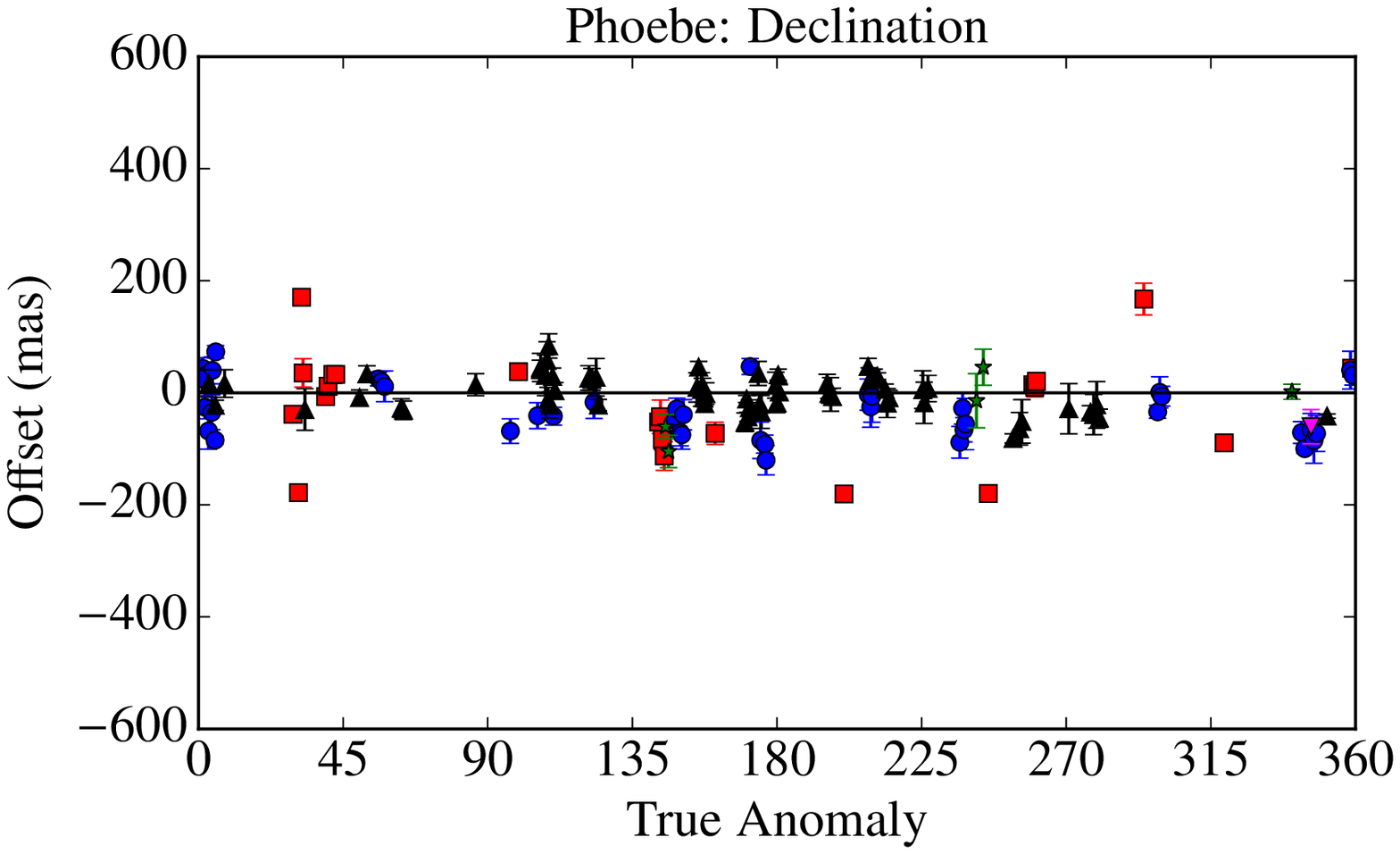}\label{Fig: an_phoebe_delta}}
\caption{Same as in Fig \ref{Fig: an_himalia_anom} for Phoebe.}
\label{Fig: an_phoebe_anom}
\end{centering}
\end{figure}

\begin{figure}[h!]
\begin{centering}
\subfigure[Right Ascension]{\includegraphics[scale=0.5]{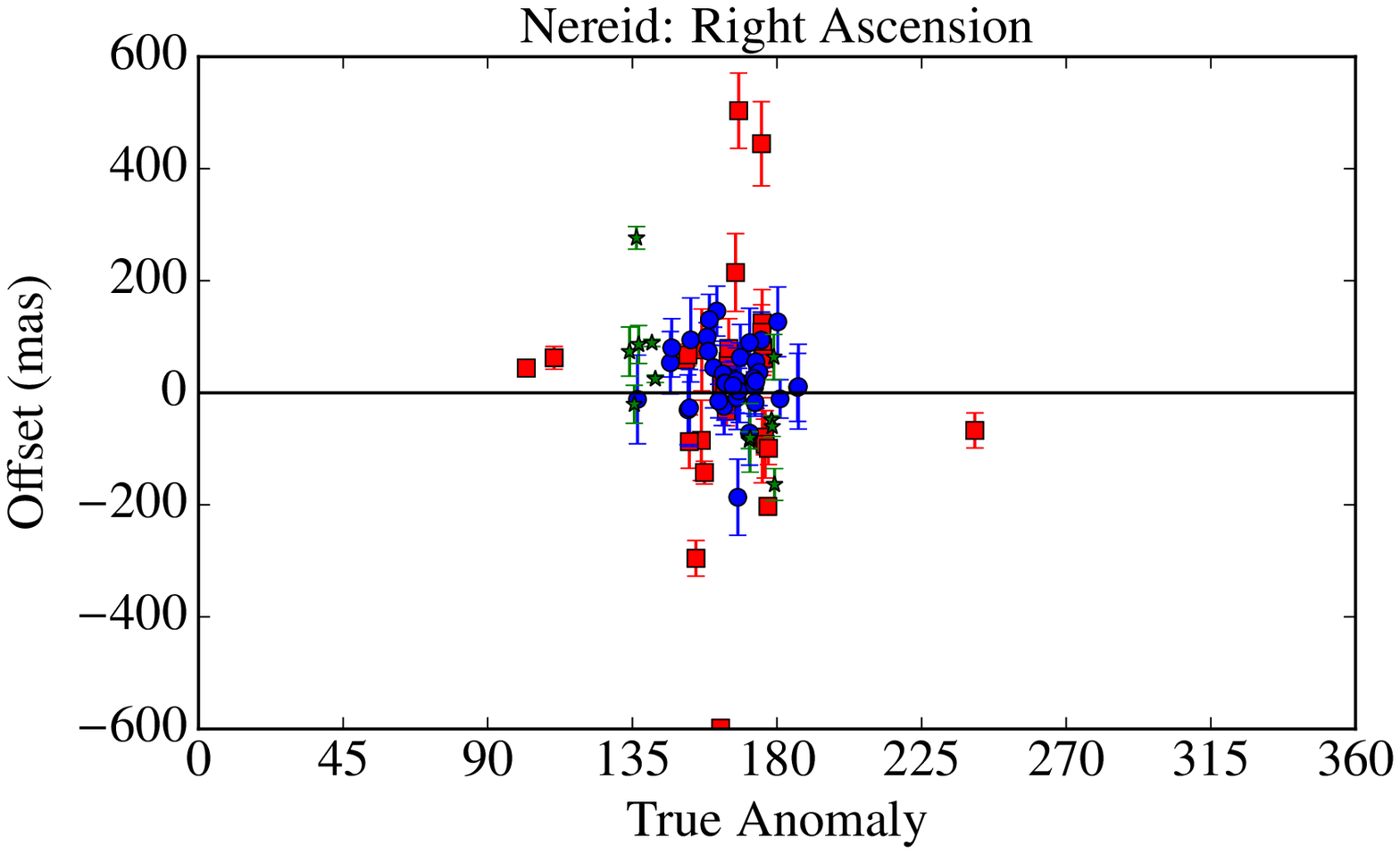}\label{Fig: an_nereid_alfa}}
\subfigure[Declination]{\includegraphics[scale=0.5]{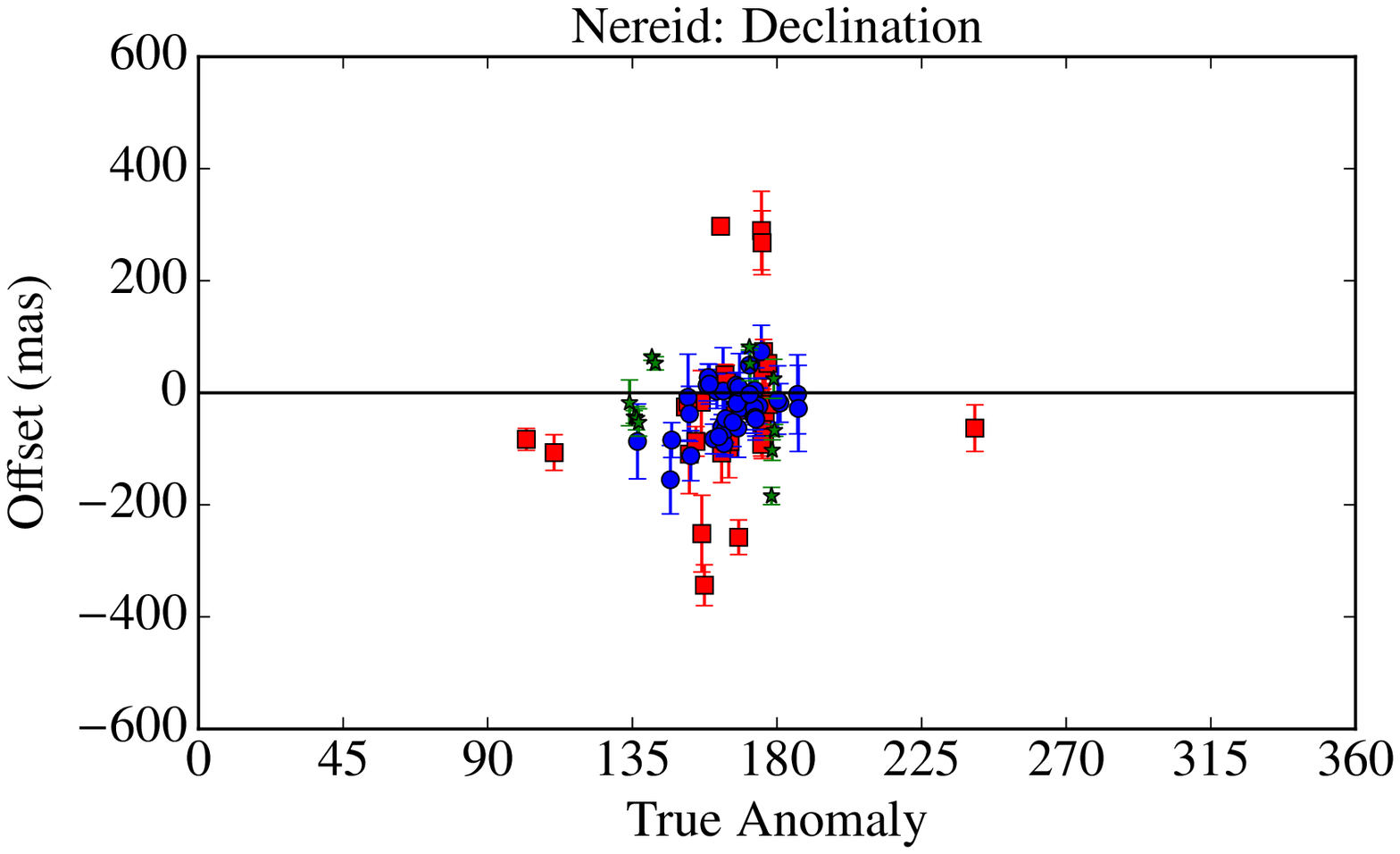}\label{Fig: an_nereid_delta}}
\caption{Same as in Fig \ref{Fig: an_himalia_anom} for Nereid.}
\label{Fig: an_nereid_anom}
\end{centering}
\end{figure}

\end{appendix}

\end{document}